\documentclass[
  a4paper,
  accepted=2026-04-18,
  amsmath,
  amssymb,
  tightenlines
]{quantumarticle}

\pdfoutput=1

\usepackage[numbers,sort&compress]{natbib}

\usepackage[utf8]{inputenc}
\usepackage[english]{babel}
\usepackage[T1]{fontenc}
\usepackage{amsmath}
\usepackage{hyperref}
\usepackage{graphicx}
\usepackage{dcolumn}
\usepackage{bm}
\usepackage{braket}
\usepackage{comment}
\usepackage{hhline}
\usepackage{float}
\usepackage{booktabs}
\usepackage{tikz}
\usepackage{quantikz}
\usepackage{xcolor}
\usepackage{dsfont}

\usepackage{ragged2e}
\usepackage{subfig}

\newcommand{\updownarrows}{\mathbin\uparrow\hspace{-.5em}\downarrow}

\definecolor{br}{RGB}{165,42,42}
\definecolor{sb}{RGB}{176,196,222}
\definecolor{crimson}{rgb}{.8, 0, 0}

\begin{document}

\title{Phase-Sensitive Measurements on a Fermi-Hubbard Quantum Processor}

\makeatletter
\def\@printtitletextwithappropriatefontsize{%
  \begingroup
    \def\@title{Phase-Sensitive Measurements on a\\Fermi--Hubbard Quantum Processor}%
    \@titleatfontsize{\huge}%
  \endgroup
}
\makeatother

\author{Alberto R.~Cavallar}
\affiliation{Max-Planck-Institut für Quantenoptik, Hans-Kopfermann-Str.~1, D-85748 Garching, Germany}
\affiliation{Munich Center for Quantum Science and Technology (MCQST), Schellingstr.~4, D-80799 München, Germany}
\affiliation{Department of Physics and Arnold Sommerfeld Center for Theoretical Physics (ASC), Ludwig-Maximilians-Universität München, Theresienstr.~37, München D-80333, Germany}
\email{A.Cavallar@lmu.de}

\author{Luis Escalera-Moreno}
\affiliation{Hamburg University of Technology, Institute for Quantum Inspired and Quantum Optimization, Blohmstraße 15, Hamburg, Germany}
\affiliation{Max-Planck-Institut für Quantenoptik, Hans-Kopfermann-Str.~1, D-85748 Garching, Germany}
\affiliation{Munich Center for Quantum Science and Technology (MCQST), Schellingstr.~4, D-80799 München, Germany}

\author{Titus Franz}
\affiliation{Max-Planck-Institut für Quantenoptik, Hans-Kopfermann-Str.~1, D-85748 Garching, Germany}
\affiliation{Munich Center for Quantum Science and Technology (MCQST), Schellingstr.~4, D-80799 München, Germany}
\author{Timon~Hilker}
\affiliation{Department of Physics and SUPA, University of Strathclyde, Glasgow, G4 0NG, United Kingdom}
\affiliation{Max-Planck-Institut für Quantenoptik, Hans-Kopfermann-Str.~1, D-85748 Garching, Germany}
\affiliation{Munich Center for Quantum Science and Technology (MCQST), Schellingstr.~4, D-80799 München, Germany}
\author{J.~Ignacio~Cirac}
\author{Philipp M.~Preiss}
\affiliation{Max-Planck-Institut für Quantenoptik, Hans-Kopfermann-Str.~1, D-85748 Garching, Germany}
\affiliation{Munich Center for Quantum Science and Technology (MCQST), Schellingstr.~4, D-80799 München, Germany}

\author{Benjamin F.~Schiffer}
\email{Benjamin.Schiffer@mpq.mpg.de}

\affiliation{Max-Planck-Institut für Quantenoptik, Hans-Kopfermann-Str.~1, D-85748 Garching, Germany}
\affiliation{Munich Center for Quantum Science and Technology (MCQST), Schellingstr.~4, D-80799 München, Germany}

\maketitle

\begin{abstract}
Fermionic quantum processors are a promising platform for quantum simulation of correlated fermionic matter.
In this work, we study a hardware-efficient protocol for measuring complex expectation values of the time-evolution operator, commonly referred to as Loschmidt echoes, with fermions in an optical superlattice.
We analyze the algorithm for the Fermi–Hubbard model at half-filling as well as at finite doping.
The method relies on global quench dynamics and short imaginary time evolution, the latter being realized by architecture-tailored pulse sequences starting from a product state of plaquettes.
Our numerical results show that complex-valued Loschmidt echoes can be efficiently obtained for large many-body states over a broad spectral range.
This allows one to measure spectral properties of the Fermi–Hubbard model, such as the local density of states, and paves the way for the study of finite-temperature properties in current fermionic quantum simulators. 
\end{abstract}

\section{Introduction} \label{sec:intro}

The Fermi–Hubbard model stands as a prototypical model in quantum many-body physics, describing a rich variety of physical phenomena of correlated electrons~\mbox{\cite{Lee_doping_2006, Qin_the-hubbard_2022}}.
While being one of the simplest models of fermions on a lattice, it provides insight into the electron-pairing mechanisms believed to underlie high-temperature superconductivity~\cite{Lee_doping_2006}.
Recent years have seen significant progress in probing the features of the model, with numerical methods~\cite{Qin_the-hubbard_2022} and through experimental efforts~\cite{bakr_microscopy_2025}.
Nonetheless, a full understanding of the Fermi–Hubbard model continues to be a central challenge.

Analog quantum simulators have proven highly successful in studying properties of the Fermi–Hubbard model~\cite{tarruell_quantum-simulation_2018}.
In particular, fermionic simulators avoid both the overhead of fermion-spin mappings~\cite{Havlicek_operator_2017} and from Trotterization~\cite{childs_theory_2021}.
However, the limited programmability of these analog devices generically restricts the physical information one can obtain beyond local observables.
While probing the spectrum of the Hamiltonian is a central objective in the quantum simulation of many-body systems, its inherently global character makes it inaccessible via the local measurements on existing analog platforms.

\begin{figure*}
    \centering
    \includegraphics[width=.95\linewidth]{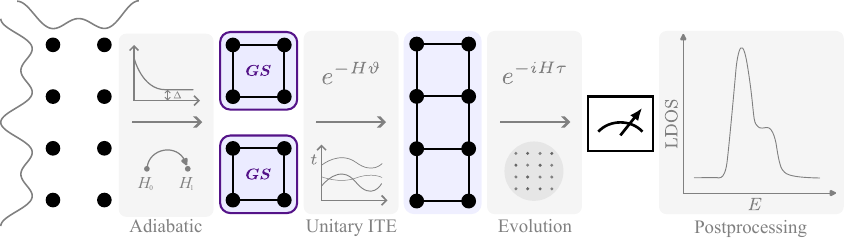}
    \caption{
    \justifying
    \textbf{Schematic overview of a control-free algorithm for probing complex-valued Loschmidt echoes on a fermionic quantum processor.}
    By leveraging an additional superlattice, we propose to prepare an initial plaquette state, with each plaquette in its adiabatically prepared ground state (GS).
    Through an optimized modulation of the hopping parameters, short imaginary time evolution (ITE) can be realized.
    Next, analog evolution under the full Fermi–Hubbard Hamiltonian is applied, followed by a projective measurement on the initial state, which is realized by reversing the adiabatic state preparation.
    Classically post-processing the measurement data and combining it with separately obtained Loschmidt echo amplitudes, yields the complex-valued Loschmidt echo as a function of the evolution time. 
    This enables computation of the local density of states (LDOS), or physical observables for microcanonical ensembles.}
    \label{fig:fig1}
\end{figure*}

A standard algorithm for estimating eigenenergies is quantum phase estimation (QPE), but it requires global time-evolution operators, which are controlled by an auxiliary qubit~\cite{kitaev_quantum_1995,nielsen_quantum_2010}. 
These interference protocols therefore involve deep quantum circuits, making the realization of large-scale QPE challenging on digital computers and unfeasible on analog simulators.
Variants of QPE exist that rely only on the expectation value of the time-evolution operator, also referred to as the \emph{complex-valued} Loschmidt echo~\cite{Lin_heisenberg_2022, blunt_statistical_2023, lu_algorithms_2021}. 
The Loschmidt echo can be interpreted as the overlap of a quantum state with a time-evolved copy of the state.
Crucially, the full Loschmidt echo requires not only the amplitude of the overlap, but also knowledge of the phase information.
While the so-called Hadamard test allows one to perform an interferometric measurement to compute the phase, it too demands controlled global dynamics~\cite{cleve_quantum_1998, nielsen_quantum_2010}.
Alternatively, the phase-sensitive measurement can be realized by creating a suitable superposition with a reference state for which the phase of the Loschmidt echo is known, thereby avoiding controlled operations.
Yet preparing these superpositions can be as challenging as a Hadamard test~\cite{obrien_error_2020, lu_algorithms_2021, hemery_measuring_2024}.
To circumvent these limitations, new hardware-efficient protocols have recently been proposed to measure the complex-valued Loschmidt echo without controlled operations or reference states~\cite{yang_phase-sensitive_2024, clinton_quantum_2024, schiffer_hardware-efficient_2025}.
These methods combine the strengths of analog platforms with additional capabilities thus offering a higher degree of flexibility~(cf.~\cite{Bluvstein_a-quantum-processor_2022,mark_efficiently_2024,Tabares_programming_2025,Karch_probing_2025}), in what we refer to as a fermionic quantum processor.

In this work, we analyze the hardware-efficient algorithm from Ref.~\cite{yang_phase-sensitive_2024}, which relies on short imaginary time evolution (ITE), for obtaining complex-valued Loschmidt echoes on a fermionic quantum processor.
We consider a quantum simulator architecture that is able to natively implement Hamiltonian dynamics of the Fermi–Hubbard model, but also allows for operations on small subsystems via an optical superlattice~\cite{gkritsis_simulating_2024, bojovic_high-fidelity_2025}.
As initial states, we consider a tiling of small plaquettes (see Fig.~\ref{fig:fig1}), each in their respective ground states,
extended over a large many-body system.
This is in the spirit of earlier theoretical and experimental results on preparing a state with resonating valence bonds~\cite{trebst_d-wave_2006, Rey_controlled_2009, nascimbene_experimental_2012}.
We then show that short ITE can be realized by architecture-tailored optimized pulses that act non-trivially on pairs of neighboring plaquettes. 
Combining these local pulses realizes short ITE on the global system.
Importantly, we then quench the system with the global Hamiltonian for different times, which will allow us to access spectral information of the full Fermi–Hubbard model in the algorithm.
Our results outline all necessary steps to measure complex-valued Loschmidt echoes in fermionic quantum simulators for large systems (see Fig.~\ref{fig:fig1}).

This paper is structured as follows. 
In Sec.~\ref{sec:fermiqp}, we outline a fermionic quantum processor platform based on a superlattice architecture, and review a quantum algorithm for phase-resolved measurements using the imaginary time evolution algorithm.
Then, in Sec.~\ref{sec:initial_state}, we describe the preparation of low-energy initial states as plaquette ground states via the quantum adiabatic algorithm.
Subsequently, the implementation of short imaginary time evolution on the plaquettes is described in Sec.~\ref{sec:implementation}.
After discussing the real time evolution in Sec.~\ref{sec:realtime}, we analyze the sample complexity for estimating the Fermi–Hubbard ground state energy in a numerical case study (Sec.~\ref{sec:benchmark}).
The doped case is presented in Sec.~\ref{sec:doping}, and we summarize the experimental steps of the protocol in Sec.~\ref{sec:protocol}.
We conclude by discussing the implications of our results in Sec.~\ref{sec:discussion}.

\section{Complex-valued Loschmidt echoes on a fermionic optical superlattice}
\label{sec:fermiqp}

Fermionic quantum simulators offer a hardware-efficient approach to simulate fermion dynamics without overheads due to fermion-qubit mappings~\cite{Jordan_uber_1928,Bravyi_fermionic_2002,Verstraete_mapping_2005,Havlicek_operator_2017,Yu_clifford_2025}.
Experiments with cold atoms in optical lattices can natively realize the spinful Fermi–Hubbard model on a two-dimensional lattice
\begin{align} \label{FHM}
    H_\text{FH}= -t \sum_{\langle i,j\rangle,\sigma}  (c_{i\sigma}^\dagger c_{j\sigma}^{} + \text{h.c.}) + U \sum_i n_{i\uparrow} n_{i\downarrow},
\end{align}
where $c_{i\sigma}^{\dagger}$ and $c_{i\sigma}^{}$ are fermionic creation and annihilation operators and $\sigma\in\{\uparrow,\downarrow\}$ denotes the spin.
The Hamiltonian includes nearest-neighbor hopping and on-site interaction terms.
Additional control can be realized via the combination with a superlattice that effectively isolates subsystems, as recently demonstrated~\cite{Zhang_scalable_2023,gkritsis_simulating_2024,bojovic_high-fidelity_2025,xu_neutral-atom_2025, Kiefer_protected_2025}.
Concretely, dimerizing the system into double wells allows for local operations on separate double wells, including digital gate operations~\cite{Zhang_scalable_2023,mark_efficiently_2024,Klemmer_floquet-driven_2024,zhang_observation_2025, Kiefer_protected_2025, Zhu_splitting_2025}, and with applications for quantum chemistry~\cite{gkritsis_simulating_2024}.
Hence, combining analog operations with additional capabilities such as digital gates and local pulses in such a Fermi–Hubbard quantum processor opens new possibilities to probe key properties of the many-body model.

We propose a measurement scheme for \emph{complex} Loschmidt echoes on a fermionic superlattice architecture. 
At the core of our proposal is the choice of the initial state as composed of plaquettes that are tiled to form a large many-body state. 
Importantly, this allows for entanglement within each plaquette but not between them.
Spectral properties of the full Fermi–Hubbard model can be probed by subsequently quenching the system under the global Hamiltonian.
We will refer to the state built from many plaquettes as the initial plaquette state $\ket{\Psi_p}$.
This approach is motivated by two main considerations:
First, high-fidelity state preparation is experimentally feasible on plaquettes, while the preparation of highly entangled states can be challenging.
Second, the initial plaquette state enables the adaptation of a recent control-free protocol that relies on short imaginary time evolution to obtain the phase information of the Loschmidt echo~\cite{yang_phase-sensitive_2024}.
We comment later how the protocol can be extended towards more complex initial states, e.g., by including auxiliary degrees of freedom (Sec.~\ref{sec:discussion}).

The complex-valued Loschmidt echo is defined as the overlap of the initial state $|\Psi_p\rangle$ with the state evolved under the Hamiltonian for a time duration $\tau$:
\begin{align} \label{LE}
    \mathcal{G}(\tau)=\langle \Psi_p|e^{-iH_\text{FH}\tau}|\Psi_p\rangle=r(\tau)e^{i\phi(\tau)}.
\end{align}
While the amplitude $r(\tau)$ can be obtained without control operations by simple projective measurements~\cite{Karch_probing_2025}, accessing the phase information $\phi(\tau)$ is significantly more challenging.
The rate of change of the Loschmidt-echo phase can be approximated by finite differences as
\begin{align} \label{CR}
    \frac{\mathrm{d}}{\mathrm{d}\tau} \phi(\tau) = \lim_{\vartheta\rightarrow0} \frac{\ln r(\tau - i\vartheta)-\ln r(\tau + i\vartheta)}{2\vartheta},
\end{align}
with $\vartheta$ the short imaginary time~\cite{yang_phase-sensitive_2024}.
Numerically integrating the phase gradient from known initial conditions ($\phi(0)=0$) then yields the phase $\phi(\tau)$ of the Loschmidt echo as a function of the time $\tau$.

Applying the ITE algorithm on the fermionic superlattice yields a protocol with three key steps executed on the quantum device.
First, we prepare low-energy states by initializing each plaquette in its respective ground state via an adiabatic evolution.
Then, a unitary realizing the (normalized) action of ITE, is applied jointly on pairs of plaquettes.
Finally, the full system evolves under the global Hamiltonian without the superlattice, before performing a projective measurement onto the initial plaquette state.
Together, this allows us to compute the rate of change $\mathrm{d}\phi(\tau)/\mathrm{d}\tau$, which after integration yields the phase of the Loschmidt echo as a function of the evolution time $\tau$.
We illustrate the main steps of the protocol in Fig.~\ref{fig:fig1} and provide details for each step in the following sections.

\begin{figure*}
    \centering
    \includegraphics[width=1\linewidth]{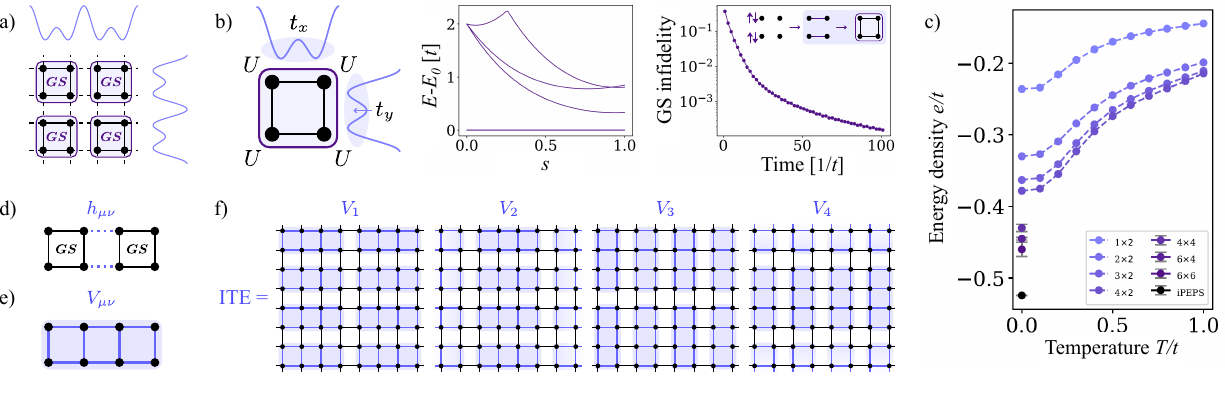}
    \caption{
    \justifying
    \textbf{Implementing short imaginary time evolution (ITE) on an initial plaquette state.}
    \textbf{(a)} An optical superlattice isolates small plaquettes, here each of size $2\times2$.
    The initial state is the product of plaquette ground states (GS). 
    \textbf{(b)} Each plaquette ground state can be prepared adiabatically from the two doublon Fock states $|\updownarrows,0\rangle$ (each on a horizontal double well).
    First, each doublon is delocalized to prepare the initial state for the second adiabatic sweep.
    Then, the plaquette ground state is prepared by slowly increasing the hopping along the vertical direction $t_y\rightarrow t$, jointly with the on-site interaction $U\rightarrow 8\,t$, while leaving $t_x =t$ fixed~(see also Ref.~\cite{nascimbene_experimental_2012}). 
    This allows for high-fidelity state preparation within a few tunneling times (few ~$1/t$).
    \textbf{(c)} Mean energy density $e/t$ of the initial plaquette ground state and at thermal equilibrium at a dimensionless temperature $T/t$ ($k_B\!=\!1$). 
    Growing the plaquette size, the energy density of the plaquette ground state approaches the ground state energy density of the full system in the thermodynamic limit; iPEPS result from Ref.~\cite{Corboz_improved_2016}.
    \textbf{(d)} For short ITE and the plaquette ground state, only ITE for the hopping terms (e.g., $h_{\mu\nu}$) across two plaquettes $P_\mu$ and $P_\nu$ needs to be implemented. 
    \textbf{(e)} The ITE is realized (up to normalization) by a unitary $V_{\mu\nu}$ acting jointly on two plaquettes. 
    \textbf{(f)} The unitary is used as building block to realize ITE on the full system by combining it into the four layers $V_1,\dots ,V_4$.}
    \label{fig:fig2}
\end{figure*}

\section{Preparation and energy of initial plaquette states} \label{sec:initial_state}

We consider initial plaquette states $\ket{\Psi_p}$, which consist of a plaquette unit cell that is repeated over the lattice, as indicated in Fig.~\ref{fig:fig2}(a).
To prepare low-energy initial states for the two-dimensional Fermi–Hubbard model at half-filling, we choose each plaquette in its respective ground state.
For concreteness, we consider the strong-coupling regime $U/t=8$ and plaquettes of size~$2\times 2$, as in Fig.~\ref{fig:fig2}(a), which allows for simple manipulation with an optical superlattice.
The plaquettes can be prepared in their ground state using a two-step adiabatic protocol on the small subsystem~\cite{farhi_quantum_2000}.
The feasibility of such a protocol has been demonstrated for bosons in Ref.~\cite{nascimbene_experimental_2012}.
We outline the adiabatic preparation protocol assuming each plaquette is initialized as two isolated horizontal double wells, each in the Fock state $|\updownarrows,0\rangle$. These are then delocalized, followed by ramping up the vertical hopping $t_y\rightarrow t$ and on-site interaction $U\rightarrow 8\,t$ while keeping the horizontal hopping $t_x=t$ fixed (see Appendix~\ref{appendix:adiabatic_prep} for details).
Note that the adiabatic preparation of the ground state $\ket{\psi_1}$ of the plaquette Hamiltonian $H_{1} =H_{\text{FH}}^{2\times 2}(t, U\!=\!8\,t)$ can be realized in parallel on all plaquettes to prepare the initial plaquette state for the ITE algorithm $\ket{\Psi_p}=\otimes_\mu \ket{\psi_1}_\mu$.
We illustrate the adiabatic protocol and include the low-energy spectrum along the adiabatic path in Fig.~\ref{fig:fig2}(b), together with the ground state infidelity as a function of the total sweep time.
For a typical hopping strength of up to 1\,kHz, 
the preparation is much faster than typical timescales of atom loss or technical heating in cold-atom platforms~\cite{bojovic_high-fidelity_2025,xu_neutral-atom_2025}.
We therefore expect the adiabatic state preparation to be robust against these sources of error.

Next, we comment on the mean energy of the initial plaquette state. 
The mean energy is a relevant indicator for the energies --- and thus temperatures --- of the Fermi–Hubbard Hamiltonian that can be reached with our ansatz state.
We first note that the energy density 
\begin{align}
    e=E/N= \langle\Psi_p|H_\text{FH}|\Psi_p\rangle/N
\end{align}
of the initial plaquette state $\ket{\Psi_p}$ evaluated over the homogeneous Fermi–Hubbard Hamiltonian $H_\text{FH}$ corresponds to the energy density of a single plaquette.
To see this, we define the Hamiltonian describing the local hopping terms between two neighboring plaquettes $P_\mu$ and $P_\nu$ as
\begin{align} \label{hopp}
    h_{\mu\nu}= -t \sum_\sigma \sum_{\langle i,j\rangle\in \Lambda_{\mu\nu}}  (c_{i\sigma}^\dagger c_{j\sigma} + \text{h.c.}),
\end{align}
with $\Lambda_{\mu\nu}$ the set of edges directly connecting sites from~$P_\mu$ with sites from~$P_\nu$.
For the initial plaquette state $\ket{\Psi_p}$, there are no energy contributions from these hopping terms between decoupled plaquettes due to orthogonality between the initial and hopped configurations, i.e.~$\langle \Psi_p|h_{\mu\nu}|\Psi_p\rangle = 0$, as every single plaquette has a fixed particle number for each spin. 
Therefore, we can compute the energy density of the Fermi–Hubbard model at small system sizes to directly obtain the energy density of the initial plaquette state at different plaquette sizes.
In Fig.~\ref{fig:fig2}(c) we show the mean energy density for increasingly larger plaquette states.
The ground state data is obtained using tensor network methods, and for small system sizes the data is complemented with energy values for the state at thermal equilibrium $\rho \sim \exp(-H\beta)$ at an inverse temperature $\beta = (k_BT)^{-1}$ using exact diagonalization. 
We set $k_B=1$ in the numerical simulations; additional details are included in Appendix~\ref{appendix:TN_data}.
From the data shown, we obtain an estimate for the energy regime accessible with a certain plaquette size.
If the mean energy of the target state is far away from the energy of the initial plaquette state, the overlap between them will typically be reduced, which can make the sample cost of the protocol prohibitively high when considering applications in phase estimation (see also Sec.~\ref{sec:benchmark} in this context).
This applies to both textbook phase estimation~\cite{nielsen_quantum_2010} and methods based on the complex-valued Loschmidt echo.
In our method, we additionally require that the amplitude of the Loschmidt echo does not become too small along the relevant time window, since the phase is obtained by integration of Eq.~\eqref{CR}. 
We refer to Ref.~\cite{schiffer_hardware-efficient_2025} for an extended discussion of the sample complexity.

Beyond phase estimation, access to the Loschmidt echoes also allows one to access observables of thermal equilibrium states when additionally performing sampling of initial states~\cite{lu_algorithms_2021}.
The mean energy of a thermal state at a fixed system size lies above the mean energy of the ground state, therefore one may use the data in Fig.~\ref{fig:fig2}(c) also to assess which temperature range is accessible with the initial plaquette states.
For the scope of this work, we focus on ground state energy estimation, however, one may also target the local density of states at higher energies with this method.

\section{Optimized pulses for imaginary time evolution} \label{sec:implementation}

In our protocol, to compute the phase information of the Loschmidt echo, we measure the amplitudes 
\begin{align} \label{shifted}
    r(\tau\pm i\vartheta)=|\langle{\Psi_p}|e^{-i H_\text{FH}\tau} e^{\pm H_\text{FH}\vartheta}|{\Psi_p}\rangle|
\end{align}
for a \textit{short} imaginary time $\vartheta$ with respect to one tunneling time unit. 
With Eq.~\eqref{CR} for finite $\vartheta$, this allows us to estimate the rate of change of the phase, $\mathrm{d}\phi(\tau)/\mathrm{d}\tau$, up to $\mathcal{O}(\vartheta^2)$ by a finite difference, which is then integrated.
We briefly recall the main steps of the protocol that are executed on the quantum device:
to measure $r(\tau\pm i\vartheta)$, one prepares the initial plaquette state $\ket{\Psi_p}$ starting from a Fock state by an adiabatic evolution $\mathcal{U}_{0\rightarrow 1}$.
Then, we require the implementation of short ITE, to prepare a state proportional to $\exp(\pm H_\text{FH}\vartheta)|{\Psi_p}\rangle$.
As a next step on the device, we perform quench dynamics with the full Fermi–Hubbard Hamiltonian, $\exp(-i H_\text{FH}\tau)$.
Finally, we project onto the initial state $\ket{\Psi_p}$.
This projection can be performed by undoing the state preparation, i.e., reversing the adiabatic evolution as $\mathcal{U}_{1\rightarrow 0}$ (see Appendix~\ref{appendix:proj_adiabatic} for details) and measuring in the Fock basis using a quantum gas microscope to reveal the overlap with the initial state~\cite{Bakr_a-quantum-gas_2009,Parsons_site-resolved_2015,Cheuk_quantum-gas_2015}.

In principle, ITE for a known initial state can always be realized by finding the unitary operation that maps the initial state to the (normalized) state after ITE.
However, this requires exactly diagonalizing the system and becomes intractable already for several sites.
As we require ITE for a short imaginary time $\vartheta$ only, we perform a first-order Trotter decomposition of the Hamiltonian $H_\text{FH}$ into a single circuit layer.
Together with the structure of the initial plaquette state $\ket{\Psi_p}$, we observe that the system factorizes into unentangled subsystems.
This allows one to restrict the task of finding unitaries $V_m$ implementing ITE to each small subsystem.
We require that the $V_m$ locally implements the ITE for a Hamiltonian term $H_m$, such that
\begin{align} \label{ITEmpsi}
    \frac{1}{\mathcal{N}_m^\pm}e^{\pm H_m\vartheta}|\Psi_p\rangle \overset{!}{=} V^\pm_m |\Psi_p\rangle,
\end{align}
with a normalization factor $\mathcal{N}_m^\pm = ||\exp(\pm H_m\vartheta)|{\Psi_p}\rangle||$.
Jointly, all ITE-unitaries then realize $\prod_m V_m^\pm \ket{\Psi_p} \propto \exp(\pm H_\text{FH}\vartheta)\ket{\Psi_p}$.
We remark that the unitaries $V_m^\pm$ depend on the sign of the ITE and also on the initial state $\ket{\Psi_p}$.

Next, we outline how the choice of the initial plaquette state $\ket{\Psi_p}$ allows for a particularly hardware-efficient ITE implementation.
Concretely, each local term $H_m$ will correspond to a pair of local hopping terms $h_{\mu\nu}$ [cf.~Eq.~\eqref{hopp}], while the remaining hopping and on-site interaction terms in Eq.~\eqref{FHM} will not contribute.
This can be understood by recalling that $h_{\mu\nu}$ contains the two hopping terms for each spin degree of freedom between a plaquette $P_\mu$ and $P_\nu$. 
Similarly, we define $h_{\mu\mu}$ as the hoppings within $P_\mu$, and $h_\mu$ the on-site interaction terms for the four sites of $P_\mu$.
Then, considering all plaquettes in the lattice, 
$H_{\Lambda}=\sum_{\mu,\nu} h_{\mu\nu}$ collects all inter-plaquette hoppings and 
$H_\square=\sum_\mu h_{\mu\mu} + \sum_\mu h_\mu$ describes all Hamiltonian terms acting non-trivially only within plaquettes.
This accounts for all terms in the Fermi–Hubbard Hamiltonian $H_\text{FH}=H_\square+H_{\Lambda}$.
By construction, the initial state $\ket{\Psi_p}$ is the ground state of the Hamiltonian $H_\square$, and therefore applying ITE with the Hamiltonian $H_\square$ to  $\ket{\Psi_p}$ yields a global factor $\exp(\pm\lambda_p\vartheta)$ that can be computed classically from the known ground state energy $\lambda_p$ of the initial plaquette state.
Considering a Trotter decomposition of $H_\text{FH}$,
\begin{align}
    e^{\pm H_\text{FH}\vartheta}\ket{\Psi_p} & = e^{\pm H_\Lambda\vartheta} e^{\pm H_\square\vartheta} \ket{\Psi_p} + \mathcal{O}(\vartheta^2)\nonumber\\
    & =  e^{\pm \lambda_p\vartheta} e^{\pm H_\Lambda\vartheta} \ket{\Psi_p}   + \mathcal{O}(\vartheta^2),
\end{align}
this shows that we only need to find a unitary realizing $\exp(\pm H_\Lambda \vartheta)|{\Psi_p}\rangle$ to implement the ITE.

Therefore, ITE has to be realized for each hopping term $h_{\mu\nu}$ [Eq.~(\ref{hopp})] between plaquettes, as we visualize in Fig.~\ref{fig:fig2}(d).
This operation is implemented up to normalization by a unitary $V_{\mu\nu}$ acting on the eight involved sites [Fig.~\ref{fig:fig2}(e)].
Such unitary serves as a building block to implement the ITE on the whole lattice through the sequence of four layers $V_1,...,V_4$ in Fig.~\ref{fig:fig2}(f), where the blocks within each layer commute with each other.
Given that gates in the first layer $V_1$ weakly entangle pairs of plaquettes, one may ask why each gate in the second layer $V_2$ acts on no more sites than in the previous one.
However, as we detail in the Appendix~\ref{appendix:ITE_scalability}, the resulting additional error is bounded by $\mathcal{O}(\vartheta^2)$ and therefore of the same magnitude as the Trotter error. 

\begin{figure}[tb]
    \centering
    \includegraphics[width=.9\linewidth]{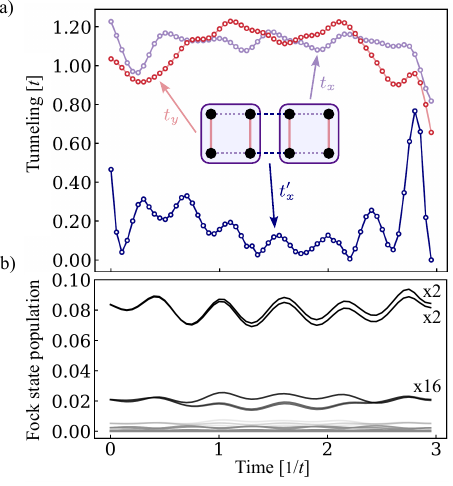}
    \caption{
    \justifying
    \textbf{Pulse sequence for imaginary time evolution. (a)} 99.9\%-fidelity pulse implementing (up to the normalization) $\exp(-H_\text{FH}\vartheta)\ket{\Psi_p}$ for $\vartheta=0.1/t$ on two half-filled plaquettes via modulation of the Fermi–Hubbard tunneling parameters $t_x, t_x', t_y$ while keeping $U=8\,t$ fixed.
    \textbf{(b)} Instantaneous Fock state populations during the pulse sequence, revealing four dominant contributions $(\approx 8\%$ each) corresponding to the possible combinations of Néel states on both plaquettes.}
    \label{fig:fig3}
\end{figure}

Next, we obtain a description of the $V_{\mu\nu}$ by finding a pulse sequence where we modulate the hopping strengths of the Fermi--Hubbard Hamiltonian on the subsystem of the plaquettes $P_\mu$ and $P_\nu$.
The task can be considered as an optimal control problem where we maximize the fidelity of preparing the (normalized) imaginary-time evolved state under the allowed operations, in the spirit of the GRAPE protocol~\cite{khaneja_optimal_2005}
as detailed in Appendix~\ref{appendix:ITE_pulse_optim}.
In the optimization, we focus on ITE with a positive imaginary time $\vartheta=0.1/t$, but the approach can be employed for other (negative) values of $\vartheta$ as well.
Following the optimal control approach, we take the hopping parameters $t_x, t_x',t_y$ as control variables while leaving $U=8\,t$ fixed.
This choice follows from the fact that both plaquettes are identical and prepared in symmetric ground states, rendering tunneling amplitudes related by plaquette and lattice symmetries equivalent.
Fig.~\ref{fig:fig3}(a) shows the optimized pulse: a smooth modulation of the three different tunneling parameters can realize ITE on the plaquette pair with a fidelity of 99.9\%.
The total pulse duration is roughly one period of the population oscillation, for an imaginary time of $\vartheta=0.1/t$. 
The independent control of hopping terms $t_x$ and $t_x'$ can be realized using the superlattice architecture.

To obtain some physical intuition of the optimized pulse sequence, we show the instantaneous projection onto the Fock state basis at different evolution times in Fig.~\ref{fig:fig3}(b).
There are four main contributions with an overlap squared of around 8\%, which partially overlap in the figure, corresponding to all possible combinations of Néel states on both plaquettes. 
The next contributions with around $2\%$ population each correspond to one of sixteen combinations of a Néel state on one plaquette and a striped (e.g.,~$\big|\begin{smallmatrix}
        \uparrow & \uparrow\\
        \downarrow & \downarrow
    \end{smallmatrix}\big\rangle$) state on the other plaquette.
We observe that final Fock state populations are similar to the initial ones, which is expected as the overall effect of short ITE is small. 
However, at intermediate times the populations oscillate, giving rise to slightly different phase factors for each Fock state.
These relative phase factors together realize the effect of the ITE.

For larger plaquette sizes it is also possible to find optimized pulse sequences with this variational approach, as long as the pair of plaquettes can be simulated classically.
Generally, pulses on larger plaquettes require individually varying the local hopping strengths, and longer pulse durations are necessary.
Alternatively, short ITE could also be decomposed into digital gate sequences, but optimizing the unitary dynamics on the pulse level allows one to find operations that achieve the same fidelity with a shorter total pulse time.

\section{Analog real time evolution} \label{sec:realtime}

After realizing the short ITE on the initial plaquette state, we implement real time evolution
$\exp(-i H_\text{FH}\tau)$ under the global Fermi–Hubbard Hamiltonian $H_\text{FH}$.
We recall that, taken together, these steps allow us to estimate the required amplitudes
$r(\tau\pm i\vartheta)$ for the ITE algorithm [cf.~Eq.~\eqref{shifted}].
The Hamiltonian dynamics are performed in an \textit{analog} mode; therefore, no additional overhead is required, e.g., due to Trotterization~\cite{childs_theory_2021}.

Up to this point, the ITE pulses and the plaquette preparation act on small, mutually decoupled subsystems and are therefore classically simulable.
By contrast, the ensuing global evolution couples distinct plaquettes and generates entanglement across the lattice, which is the step that lifts the protocol to extended many-body systems and also renders classical simulation costly.

We repeat the real time evolution for a set of total times $\{\tau_m\}$, as required by the ITE algorithm~\cite{yang_phase-sensitive_2024}.
The largest accessible time $\tau_{\max}$ sets the minimal resolvable energy scale in applications of the Loschmidt echo.
Also, the choice of the sampling grid $\{\tau_m\}$ generally depends on the application. 
We study these parameters in detail in the numerical analysis in the following section.

\section{Numerical sample complexity \mbox{analysis}} \label{sec:benchmark}

\begin{figure*}
    \centering
    \includegraphics[width=\linewidth]{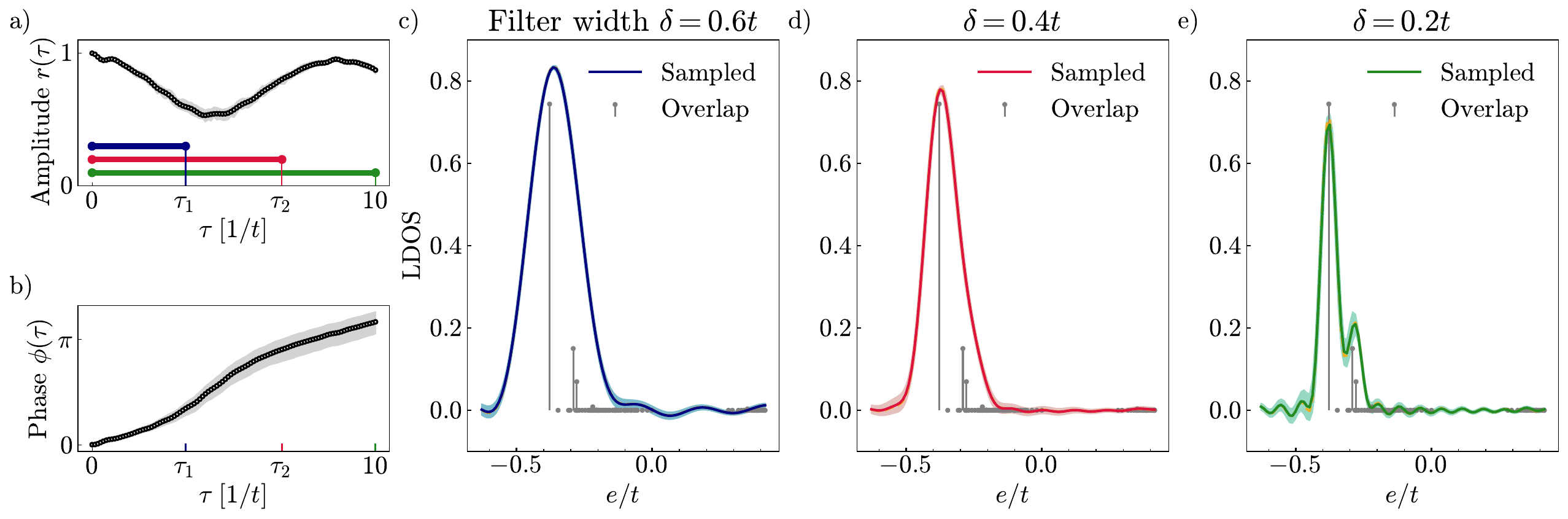}
    \caption{
    \justifying
    \textbf{Shot noise analysis of the local density of states (LDOS).}
    The amplitude \textbf{(a)} and phase \textbf{(b)} of the Loschmidt echoes are computed as a function of time for the Fermi–Hubbard Hamiltonian at $U/t=8$ on a $4\times 2$ lattice at half-filling.
    The mean-energy contribution to the phase has been subtracted for better clarity (see Appendix~\ref{appendix:numerical_samples}).
    \textbf{(c-e)} For Gaussian filter widths $\delta=0.6t,0.4t,0.2t$, Loschmidt echoes up to times $\tau_1=\tau_3/3$, $\tau_2=2\tau_3/3$, $\tau_3=10$ (in units of 
    $1/t$) 
    are respectively used to estimate the LDOS.
    Each of the amplitudes $r(\tau)$ and $r(\tau\pm i\vartheta)$ has been estimated with 100 samples. 
    The mean value is shown as a solid curve with a shaded region showing one standard deviation of the data.
    The exact curve resulting from the ITE algorithm without sampling is shown as a dashed line in yellow, hardly visible due to the high precision of the overlying sampled curves.
    Gray bars with markers indicate overlaps of the initial plaquette state 
    $\ket{\Psi_p}$ with the $4\times 2$ Fermi–Hubbard eigenstates.
    }
    \label{fig:fig4}
\end{figure*}

In our protocol, we avoid ancilla-controlled global operations to obtain the phase; however, this comes at the cost of an increased sample complexity.
The total number of samples to compute the complex-valued Loschmidt echo with the ITE algorithm up to a time $\tau$ and for a maximal additive error $\epsilon$ in each Loschmidt echo on an $N$-site system is given as
\begin{align}
    M_\text{samples}=\mathcal{O}\big(N\tau^3/\epsilon^3r_{\pm, \text{min}}^2\big)
\end{align}
where $r_{\pm, \text{min}}$ is a lower bound on the minimal Loschmidt echo amplitude during the integration window~\cite{yang_phase-sensitive_2024, schiffer_hardware-efficient_2025}.
This scaling takes the shot noise and the Trotter error of approximating the ITE into account.
Our interest extends beyond the asymptotic behavior to the concrete number of samples for a practical application.
To this end, we perform numerical simulations of the protocol for small system size, assuming exact ITE implementation and ignoring hardware imperfections.

As an application for the Loschmidt echo, we consider the local density of states~(LDOS), which describes the projection of the initial state $\ket{\Psi_p}$ on the energy eigenvalues $E_n$ of a target Hamiltonian:
\begin{align}
    D_{\Psi_p}(E)=\sum_n |\langle \Psi_p|E_n\rangle|^2 \delta(E-E_n).
\end{align}
If the initial state has non-negligible overlap with the Hamiltonian ground state, the LDOS provides the ground state energy.
The LDOS can be obtained by Fourier transforming the time-series of the complex-valued Loschmidt echo.
Considering a Gaussian energy filter $P_\delta(E) = \exp[-(H-E)^2/2\delta^2]$ of width $\delta$, the LDOS can be expressed as a sum
\begin{align} \label{eq:DOS}
    D_{\delta,\Psi_p}(E) = \langle\Psi_p|P_\delta(E)|\Psi_p\rangle \approx \!\!\!\sum_{m=-\infty}^{\infty}\!\!\! c_m e^{iE\tau_m}\mathcal{G}(\tau_m),
\end{align}
with $\mathcal{G}(\tau_m)=\langle\Psi_p|\exp(-iH_\text{FH}\tau_m)|\Psi_p\rangle$ evaluated at discrete times~$\tau_m$ and classically computed coefficients~$c_m$.
In practice, the Loschmidt echo is only available until a finite time.
This sets an effective limit on the available filter width $\delta$ via the well-known trade-off between the total time and the desired frequency resolution when performing a Fourier transformation, yielding a relation $\tau_\text{total}\sim 1/\delta$.

Concretely, we estimate the local density of states~(LDOS) for the Fermi–Hubbard model on a~$4\times 2$-lattice using Loschmidt echoes. 
The initial plaquette state $\ket{\Psi_p}$ is composed of two $2\times2$-plaquettes, as described in the previous sections.
In Figs.~\ref{fig:fig4}(a,b) we show the amplitude $r(\tau)$ and phase $\phi(\tau)$ of the Loschmidt echo, obtained using the ITE algorithm with a resolution of $\Delta\tau =0.1/t$, which we numerically find to yield sufficient accuracy for the LDOS.
We then use this data to compute the LDOS at different filter widths, estimated from Loschmidt echoes up to different maximum times $\tau_1, \tau_2, \tau_3$, inversely proportional to the respective filter width $\delta$.
These curves of the broadened LDOS are shown in Figs.~\ref{fig:fig4}(c-e), next to the exact overlaps of the initial plaquette state with the Fermi–Hubbard eigenstates marked with gray bars.
One finds the first peak of the LDOS around the position of the ground state energy.
We include an additional analysis where the filter width is kept constant while the maximum time is changed in Appendix~\ref{appendix:numerical_samples}.

For all scenarios, we analyze the sample complexity to estimate the LDOS from the amplitudes $r(\tau)$ and $r(\tau\pm i\vartheta)$, estimated each with 100 samples per discrete time $\tau$.
This number has proven to be sufficient here, but generally depends on the application.
Together with the time step of $\Delta\tau =0.1/t$ this yields a total number of 10\,000, 20\,000 and 30\,000 samples for the three cases respectively.
We compare the data obtained with a finite amount of samples (solid lines) to the exact one in Figs.~\ref{fig:fig4}(c-e), and find good agreement.
We estimate that the data shown in Fig.~\ref{fig:fig4} can be obtained in a few hours on next-generation fermionic quantum simulators with targeted experimental cycle times of around one second.
The data acquisition time could be further reduced by performing the experiment on several systems in parallel using a single optical lattice.

In practice, decoherence mechanisms such as photon-scattering from lattice or control beams lead to an effective damping of the measured amplitudes.
As analyzed in related phase-estimation protocols, such damping primarily results in an additional broadening of the local density of states, reducing the achievable precision for a fixed measurement budget but not necessarily biasing the mean value of the estimated observables (see Appendix~I in Ref.~\cite{schiffer_hardware-efficient_2025}).

While we focused the study of the sample complexity on the LDOS as a quantum phase estimation protocol, access to the complex-valued Loschmidt echo allows for much broader applications.
By generalizing the Loschmidt echo to include an observable~$O$, such as $\langle \Psi|Oe^{-iH\tau}|\Psi\rangle$, one can probe the observable for the initial state filtered at a particular energy.
These generalized Loschmidt echoes can be obtained with very little additional overhead. 
Concretely, if $O$ is described by a local unitary, e.g.,~the single-spin parity $\exp(i\pi n_j)$, it can directly be applied in the circuit. 
Otherwise, for general non-unitary observables it is always possible to decompose them into a linear combination of unitaries~\cite{childs_hamiltonian_2012}.
While the filtered state describes a microcanonical ensemble in the limit of a very narrow filter, by additionally sampling according to the Boltzmann distribution, expectation values at different equilibrium temperatures can be probed~\cite{lu_algorithms_2021}.

We remark that our method can be adapted to act as a projection method onto the eigenstates of a larger family of Hamiltonians that can be implemented in the superlattice architecture.
One example is to consider only the kinetic part of the Fermi–Hubbard Hamiltonian $H_\text{FH}(t,0) = -t \sum_{\langle i,j\rangle, \sigma} (c_{i\sigma}^\dagger c_{j\sigma} + \text{h.c.})$.
We then require pulses implementing the corresponding ITE and the on-site interaction is set to zero during the Hamiltonian evolution.

\section{Extension to the doped case} \label{sec:doping}

In the previous sections, we outlined the protocol for measuring Loschmidt echoes for the Fermi–Hubbard Hamiltonian at half-filling.
This allows for the conceptually simplest presentation of the protocol.
However, our methods are not restricted to the half-filled case and naturally extend to the doped model.
Including doping is of particular interest for studying physical phenomena of strongly correlated fermions.

\begin{figure}[tb]
    \centering
    \includegraphics[width=.9\linewidth]{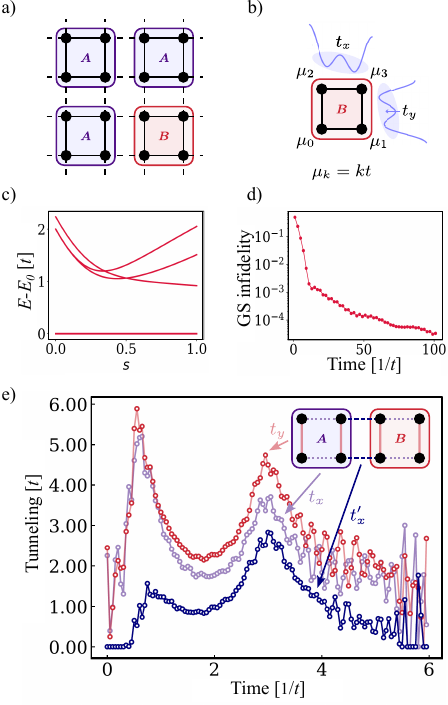}
    \caption{
    \justifying
    \textbf{Doped case.}
    \textbf{(a)} A doped initial plaquette state realized by replacing every fourth half-filled plaquette~$A$ by a doped one $B$. 
    The $B$ plaquette has one up and one down spin only, resulting in 12.5\% doping.
    \textbf{(b)} Adiabatic preparation of the doped $B$-plaquette by adding a degeneracy-breaking, spin-independent potential gradient on top of the $t_y$ and $U$ ramps. 
    \textbf{(c)} Adiabatic spectrum for the $B$-plaquette.
    \textbf{(d)} A linear sweep prepares the $B$-plaquette ground state with very high fidelity within few hopping periods. 
    \textbf{(e)} A 99.9\%-fidelity ITE pulse sequence for the $AB$ plaquette pair realized via modulation of the tunneling parameters $t_x, t_x', t_y$ while keeping $U=8\,t$ fixed.
    }
    \label{fig:fig5}
\end{figure}

We extend the plaquette ansatz to the doped case by combining plaquettes of different total spin occupation.
We consider symmetric doping of 12.5\% --- corresponding to a filling fraction of $n=0.875$ --- implying that out of every 8 up and 8 down spins, one spin from each species is removed compared to half-filling.
Concretely, we achieve this by combining three half-filled plaquettes $A$ with a fourth doped plaquette $B$ consisting of 1 up and 1 down spin only, as shown in Fig.~\ref{fig:fig5}(a).
This $AAAB$ combination has a mean energy density of $e/t=-0.57$, 
compared with the equally doped ground state energy density on a $4\times 4$ system of $e/t\approx-0.64$ (see Fig.~\ref{fig:app_dmrg_halffilled}, Appendix~\ref{appendix:TN_data}).
For reference, the iPEPS result in the thermodynamic limit is $e/t\approx-0.76$~\cite{Corboz_improved_2016}.

The doped plaquette $B$ can be adiabatically prepared, similarly to the $A$ plaquette, by introducing an additional spin-independent potential gradient $\mu_k = kt$, as illustrated in Fig.~\ref{fig:fig5}(b).
The potential gradient is necessary to break the existing degeneracies of the ground state for the doped case~\footnote{There exist other choices to break the degeneracy, such as $\mu_k=\delta_{k,0}t+\delta_{k,1}t$, but they might be more difficult to implement experimentally.}.
Then, throughout the adiabatic path, there is a finite energy gap between the ground state and the first excited state [Fig.~\ref{fig:fig5}(c)]. 
We obtain fidelities beyond 99.9\% within few hopping periods with this ansatz, as shown in Fig.~\ref{fig:fig5}(d).

Implementing short ITE on the doped state with the unit cell $AAAB$ requires an additional pulse sequence for the plaquette pairs $AB$, besides the pulse shown in Fig.~\ref{fig:fig3} for all other $AA$ plaquette pairs.
Following the same procedure, we find the optimized pulse in Fig.~\ref{fig:fig5}(e), showing that one can reach 99.9\% fidelities for the doped case (see also Appendix~\ref{appendix:Fock_doped}).
In principle, one could expect that different local tunneling amplitudes are necessary to find a pulse sequence. 
We highlight that the pulse sequence shown does not rely on additional optimization parameters, yet we are still able to find a high-fidelity pulse.
The pulse sequence may be further optimized by explicitly constraining the rate of change of system parameters to enforce smoothness and enhance robustness against errors.
While the convergence of such constrained optimizations is not generally guaranteed, we expect it to be achievable when allowing for sufficiently long pulse durations; a systematic exploration of this is left for future work.

As an alternative to the $AAAB$ configuration, one can combine two half-filled plaquettes $A$ with a doped plaquette $C$ missing a spin-down particle and a plaquette $D$ missing a spin-up.
However, this $ACAD$ combination is at a mean energy density of $e/t=-0.46$, i.e.,~at higher energy than the $AAAB$ configuration.
We include further details on this variant in Appendix~\ref{appendix:alternative_doping}.
Together, we demonstrate that the plaquette ansatz for computing Loschmidt echoes allows to be naturally extended to the doped model.

\section{Summary of the experimental \mbox{protocol}} \label{sec:protocol}

We include a summary of the experimental protocol to measure \emph{complex-valued} Loschmidt echoes on a Fermi–Hubbard quantum processor:

\begin{enumerate}
    \item Prepare the initial plaquette ground states:
    \begin{enumerate}
         \item Initialize the Fock state consisting of doublons $|\updownarrows\rangle$ on every second site. 
         For the doped case, remove a doublon on every fourth plaquette. 
        \item Adiabatically prepare the plaquette ground state.
    \end{enumerate}
    \item Implement short ITE by applying the unitaries $\prod_{m=1}^4 V_m$, from Fig.~\ref{fig:fig2}(f), each described by the pulse sequence in Fig.~\ref{fig:fig3}(a). 
    Employ corresponding pulses for the doped plaquettes. 
    \item Time-evolve the full system for a duration $\tau$ in the analog mode.
    \item For the measuring step: Reverse the adiabatic ramps (see~Appendix~\ref{appendix:proj_adiabatic}) to uncompute the state preparation.
    Measure in the Fock basis by taking a snapshot of the state and count the occurrences of the initial Fock state.
    \item Repeat all previous steps for each time step $\tau$.
    The maximum time corresponds to a minimally possible filter width (cf.~Fig.~\ref{fig:fig4}). Step 2 is only included when estimating the amplitudes $r(\tau\pm i\vartheta)$, but not for the $r(\tau)$.
    \item Finally, perform classical post-processing of the collected data:
    \begin{enumerate}
        \item Estimate the survival probabilities $r(\tau)^2$ and $r(\tau\pm i \vartheta)^2$ from the sampled data.
        \item Include normalization factors [cf.~Eq.~\eqref{ITEmpsi}], which can be computed numerically.
        \item Estimate the phase $\phi(\tau)$ of each Loschmidt echo from the amplitudes $r(\tau\pm i\vartheta)$ following Eq.~(\ref{CR}) and the ensuing numerical integration.
        \item Reconstruct the complex-valued Loschmidt echo by combining the phase $\phi (\tau)$ with the separately estimated amplitude $r(\tau)$.
    \end{enumerate}
\end{enumerate}

The hardware requirements for the protocol are already available in experiments of current fermionic superlattices~\cite{xu_neutral-atom_2025,bojovic_high-fidelity_2025}.

\section{Discussion and Outlook} \label{sec:discussion}

We have presented a hardware-efficient approach to measure complex-valued Loschmidt echoes for the Fermi–Hubbard model on a fermionic superlattice.
As initial states we consider plaquette ground states, which can be prepared adiabatically, and show that the necessary imaginary time evolution can be realized on these plaquettes with optimized pulses.
Our approach builds on the strengths of analog quantum simulators that implement time evolution without additional overheads from Trotterization~\cite{childs_theory_2021}. 
Together, this allows for phase-resolved measurements of the Fermi–Hubbard model both at half-filling and when including doping.

We focused on initial plaquette states, which allows for a particularly hardware-efficient protocol to compute phases of Loschmidt echoes. 
Our approach is scalable in the sense that the plaquettes can be easily tiled towards large systems.
Also the unitaries that act jointly on two plaquettes can be combined to implement ITE on a large initial plaquette state. 
In principle, the number of plaquettes combined in the system is only limited by the optical lattice size and the sample complexity, which increases linearly with the system size~\cite{yang_phase-sensitive_2024, schiffer_hardware-efficient_2025}.
However, we comment on a limitation of this approach when scaling to large systems: the classical cost of finding an optimized pulse, or circuit decomposition, for the ITE generally grows exponentially with the plaquette size.
In many cases, though, one is interested in computing the expectation value of an intensive quantity in the thermodynamic limit up to a constant accuracy (see also Ref.~\cite{Trivedi_quantum_2024}). 
Then, one does not require arbitrarily large plaquette sizes, but the size of the plaquette is fixed by the desired accuracy.
It is worth emphasizing that our protocol is not only applicable to measure ground state energies, but also applies to initial states with a finite energy density. 
Then, even when eigenstates cannot be individually resolved, due to very small energy gaps at finite energy, one can nevertheless probe the local density of states at different energies which effectively aggregates multiple eigenstates in the target energy window.
Accessing the strongly correlated regime of the Fermi–Hubbard model at very low energies remains challenging as it likely requires larger plaquette sizes, yet, our proposal provides the tools to extract various physical properties --- from the local density of states to finite-temperature expectation values --- in currently emerging fermionic quantum processors.

We demonstrate that it is possible to find hardware-native pulse sequences that realize ITE on the plaquettes. 
Thereby, we restrict the allowed parameter range of the tunneling to values that can be implemented experimentally.
We provide the optimization code in~\cite{ITEcode}, which may be readily adapted for different goals: to make the pulse more robust to perturbations (cf.~Ref.~\cite{mark_efficiently_2024}), to better suppress band excitations, or to account for limited bandwidth of control electronics.
If the classical computational effort to find a unitary representation of the ITE becomes prohibitively large, another avenue is to go beyond initial plaquette states and consider arbitrarily entangled initial states.
This generalization allows one to increase the overlap of the initial state with low-energy eigenstates of the Fermi–Hubbard Hamiltonian, but requires a more involved experimental protocol using auxiliary degrees of freedom and postselection, as prescribed in Ref.~\cite{schiffer_hardware-efficient_2025}.
The necessary (locally) controlled operations can be realized on a fermionic quantum processor through auxiliary fermionic spins instead of qubits.
Effectively, this constitutes an effective trade-off between a more restrictive class of initial states with a simpler experimental protocol, or a method for arbitrary initial states with additional practical challenges.

The methods presented in this paper show a practical application of the ITE algorithm~\cite{yang_phase-sensitive_2024} to complex-valued Loschmidt echoes on Fermi–Hubbard quantum simulators.
The amplitude of the Loschmidt echo can already be reliably probed in current experiments on quantum simulators~\cite{Karch_probing_2025}, and our proposal has very few additional requirements to implement short ITE.
Moreover, the required superlattice has already been demonstrated in recent experiments~\cite{xu_neutral-atom_2025, bojovic_high-fidelity_2025}.
Therefore, the protocol described here could be immediately employed on current quantum hardware.
Access to complex-valued Loschmidt echoes opens up a broad range of new applications for quantum simulators: here, we use Loschmidt echoes to probe spectral properties such as the ground state energy. The methods can also be used to lessen the requirement for very accurate state preparation by realizing an effective filter on the target state, and they also enable the computation of observables at thermal equilibrium~\cite{lu_algorithms_2021}.
Together, our results contribute to advancing quantum simulation for fermionic quantum simulators beyond current capabilities.

\begin{acknowledgments}
We thank I.~Bloch, L.~Pollet, L.~Qiu, Y.~Yang, and D.~Wild for insightful discussions.
We acknowl-edge support from 
the Max Planck Society (MPG),
the Federal Ministry of Research, Technology and Space (BMFTR) via the joint project FermiQP (grant No.~13N15889 and No.~13N15890) and from  Germany's Excellence Strategy (EXC-2111-390814868). 
This work is part of the Munich Quantum Valley, which is supported by the Bavarian state government with funds from the Hightech Agenda Bayern Plus.
L.E.M.~acknowledges support from the BMFTR within the research program ``Quantum Systems'' via the joint project NOGS (grant No.~13N17144). 
T.F.~acknowledges funding from HORIZON-CL4-2022-QUANTUM-02-SGA (project 101113690, PASQuans2.1).
T.H. acknowledges funding from the European Research Council (ERC) under the European Union's Horizon Europe research and innovation program (Grant Agreement No.~101165353 --- ERC Starting Grant FOrbQ).
P.M.P.~acknowledges funding from the European Union’s Horizon 2020 research and innovation program (Grant agreement No.~948240 --- ERC Starting Grant UniRand).
\end{acknowledgments}

\bibliographystyle{quantum}
\bibliography{biblio}

@article{Lee_doping_2006,
  title = {Doping a {Mott} insulator: {Physics} of high-temperature superconductivity},
  author = {Lee, Patrick A. and Nagaosa, Naoto and Wen, Xiao-Gang},
  journal = {Rev. Mod. Phys.},
  volume = {78},
  issue = {1},
  pages = {17--85},
  numpages = {0},
  year = {2006},
  month = {Jan},
  publisher = {American Physical Society},
  doi = {10.1103/RevModPhys.78.17}
}

@article{Qin_the-hubbard_2022,
   title={The {Hubbard} {Model}: A {Computational} {Perspective}},
   volume={13},
   ISSN={1947-5462},
   DOI={10.1146/annurev-conmatphys-090921-033948},
   number={1},
   journal={Annual Review of Condensed Matter Physics},
   publisher={Annual Reviews},
   author={Qin, Mingpu and Schäfer, Thomas and Andergassen, Sabine and Corboz, Philippe and Gull, Emanuel},
   year={2022},
   month=mar, pages={275–302}
}

@misc{bakr_microscopy_2025,
  title         = {Microscopy of {Ultracold} {Fermions} in {Optical} {Lattices}},
  author        = {Bakr, Waseem S. and Ba, Zengli and Prichard, Max L.},
  year          = {2025},
  note          = {\href{https://doi.org/10.48550/arXiv.2507.04042}{arXiv:2507.04042 [cond-mat.quant-gas]}}
}

@article{tarruell_quantum-simulation_2018,
  author    = {Leticia Tarruell and Laurent Sanchez-Palencia},
  title     = {Quantum simulation of the {Hubbard} model with ultracold fermions in optical lattices},
  journal   = {Comptes Rendus. Physique},
  volume    = {19},
  number    = {6},
  pages     = {365--393},
  year      = {2018},
  note      = {Quantum simulation / Simulation quantique},
  doi = {10.1016/j.crhy.2018.10.013}
}

@article{Havlicek_operator_2017,
   title={Operator locality in the quantum simulation of fermionic models},
   volume={95},
   ISSN={2469-9934},
   doi = {10.1103/PhysRevA.95.032332},
   number={3},
   journal={Physical Review A},
   publisher={American Physical Society (APS)},
   author={Havlíček, Vojtěch and Troyer, Matthias and Whitfield, James D.},
   year={2017},
   month=mar,
   pages = {032332}
}

@article{childs_theory_2021,
    title = {Theory of {Trotter} {Error} with {Commutator} {Scaling}},
    volume = {11},
    issn = {2160-3308},
    doi = {10.1103/PhysRevX.11.011020},
    number = {1},
    journal = {Physical Review X},
    author = {Childs, Andrew M. and Su, Yuan and Tran, Minh C. and Wiebe, Nathan and Zhu, Shuchen},
    month = feb,
    year = {2021},
    pages = {011020}
}

@article{kitaev_quantum_1995,
  title={{Quantum} measurements and the {Abelian} {Stabilizer} {Problem}},
  author={Alexei Y. Kitaev},
  journal={Electron. Colloquium Comput. Complex.~},
  year={1995},
  volume={TR96},
  doi={10.48550/arXiv.quant-ph/9511026}
}

@book{nielsen_quantum_2010,
    title = {{Quantum} {Computation} and {Quantum} {Information}},
    author = {Nielsen, Michael A. and Chuang, Isaac L.},
    year = {2010},
    publisher = {Cambridge University Press},
    doi = {10.1017/cbo9780511976667}
}

@article{Lin_heisenberg_2022,
   title={Heisenberg-{Limited} {Ground}-{State} {Energy} {Estimation} for {Early} {Fault}-{Tolerant} {Quantum} {Computers}},
   volume={3},
   ISSN={2691-3399},
   doi={10.1103/PRXQuantum.3.010318},
   number={1},
   journal={PRX Quantum},
   publisher={American Physical Society (APS)},
   author={Lin, Lin and Tong, Yu},
   year={2022},
   month=feb,
   pages = {010318}
}

@article{blunt_statistical_2023,
	title = {Statistical {Phase} {Estimation} and {Error} {Mitigation} on a {Superconducting} {Quantum} {Processor}},
	volume = {4},
	  doi = {10.1103/PRXQuantum.4.040341},
	number = {4},
	urldate = {2025-02-11},
	journal = {PRX Quantum},
	author = {Blunt, Nick S. and Caune, Laura and Izsák, Róbert and Campbell, Earl T. and Holzmann, Nicole},
	month = dec,
	year = {2023},
    pages = {040341}
}

@article{lu_algorithms_2021,
	title = {Algorithms for {Quantum} {Simulation} at {Finite} {Energies}},
	volume = {2},
	doi = {10.1103/PRXQuantum.2.020321},
	number = {2},
	journal = {PRX Quantum},
	author = {Lu, Sirui and Bañuls, Mari Carmen and Cirac, J. Ignacio},
	month = may,
	year = {2021},
	pages = {020321},
}

@article{cleve_quantum_1998,
	title = {Quantum algorithms revisited},
	volume = {454},
	issn = {1471-2946},
	doi = {10.1098/rspa.1998.0164},
	number = {1969},
	journal = {Proceedings of the Royal Society of London. Series A: Mathematical, Physical and Engineering Sciences},
	author = {Cleve, R. and Ekert, A. and Macchiavello, C. and Mosca, M.},
	month = jan,
	year = {1998},
	pages = {339--354},
}

@article{obrien_error_2020,
   title={Error {Mitigation} via {Verified} {Phase} {Estimation}},
   volume={2},
   ISSN={2691-3399},
   doi={10.1103/PRXQuantum.2.020317},
   journal={PRX Quantum},
   publisher={American Physical Society (APS)},
   author={O’Brien, Thomas E. and Polla, Stefano and Rubin, Nicholas C. and Huggins, William J. and McArdle, Sam and Boixo, Sergio and McClean, Jarrod R. and Babbush, Ryan},
   year={2021},
   month=may,
   pages = {020317}
}

@article{hemery_measuring_2024,
	title = {Measuring the {Loschmidt} {Amplitude} for {Finite}-{Energy} {Properties} of the {Fermi}-{Hubbard} {Model} on an {Ion}-{Trap} {Quantum} {Computer}},
	volume = {5},
	issn = {2691-3399},
	doi = {10.1103/PRXQuantum.5.030323},
	number = {3},
	journal = {PRX Quantum},
	author = {Hémery, Kévin and Ghanem, Khaldoon and Crane, Eleanor and Campbell, Sara L. and Dreiling, Joan M. and Figgatt, Caroline and Foltz, Cameron and Gaebler, John P. and Johansen, Jacob and Mills, Michael and Moses, Steven A. and Pino, Juan M. and Ransford, Anthony and Rowe, Mary and Siegfried, Peter and others},
	month = aug,
	year = {2024},
    pages = {030323}
}

@article{yang_phase-sensitive_2024,
	title = {Phase-{Sensitive} {Quantum} {Measurement} without {Controlled} {Operations}},
	volume = {132},
	doi = {10.1103/PhysRevLett.132.220601},
	number = {22},
	journal = {Phys. Rev. Lett.},
	author = {Yang, Yilun and Christianen, Arthur and Bañuls, Mari Carmen and Wild, Dominik S. and Cirac, J. Ignacio},
	month = may,
	year = {2024},
	pages = {220601},
}

@article{clinton_quantum_2024,
  title = {{Quantum} {Phase} {Estimation} {Without} {Controlled} {Unitaries}},
  author = {Clinton, Laura and Cubitt, Toby S. and Garcia-Patron, Raul and Montanaro, Ashley and Stanisic, Stasja and Stroeks, Maarten},
  journal = {PRX Quantum},
  volume = {7},
  issue = {1},
  pages = {010345},
  numpages = {33},
  year = {2026},
  month = {Mar},
  publisher = {American Physical Society},
  doi = {10.1103/7qcr-znl2}
  }

@article{schiffer_hardware-efficient_2025,
    title={Hardware-Efficient Quantum Phase Estimation via Local Control},
   volume={6},
   ISSN={2691-3399},
   doi={10.1103/zlp8-rz9n},
   number={4},
   journal={PRX Quantum},
   publisher={American Physical Society (APS)},
   author={Schiffer, Benjamin F. and Wild, Dominik S. and Maskara, Nishad and Lukin, Mikhail D. and Cirac, J. Ignacio},
   year={2025},
   month=dec,
   pages = {040348}
}

@article{Bluvstein_a-quantum-processor_2022,
   title={A quantum processor based on coherent transport of entangled atom arrays},
   volume={604},
   ISSN={1476-4687},
   doi={10.1038/s41586-022-04592-6},
   number={7906},
   journal={Nature},
   publisher={Springer Science and Business Media LLC},
   author={Bluvstein, Dolev and Levine, Harry and Semeghini, Giulia and Wang, Tout T. and Ebadi, Sepehr and Kalinowski, Marcin and Keesling, Alexander and Maskara, Nishad and Pichler, Hannes and Greiner, Markus and Vuletić, Vladan and Lukin, Mikhail D.},
   year={2022},
   month=apr, 
   pages={451–456}
}

@article{mark_efficiently_2024,
   title={Efficiently measuring $d$-wave pairing and beyond in quantum gas microscopes},
   volume={135},
   ISSN={1079-7114},
   doi={10.1103/dqyf-kl8x},
   number={12},
   journal={Physical Review Letters},
   publisher={American Physical Society (APS)},
   author={Mark, Daniel K. and Hu, Hong-Ye and Kwan, Joyce and Kokail, Christian and Choi, Soonwon and Yelin, Susanne F.},
   year={2025},
   month=sep,
   pages = {123402}
}

@article{Tabares_programming_2025,
  title = {{Programming} {Optical}-{Lattice} {Fermi}-{Hubbard} {Quantum} {Simulators}},
  author = {Tabares, Cristian and Kokail, Christian and Zoller, Peter and Gonz\'alez-Cuadra, Daniel and Gonz\'alez-Tudela, Alejandro},
  journal = {PRX Quantum},
  volume = {6},
  issue = {3},
  pages = {030356},
  numpages = {26},
  year = {2025},
  month = {Sep},
  publisher = {American Physical Society},
  doi = {10.1103/3nx4-bnyy}
}

@misc{Karch_probing_2025,
  title  = {Probing quantum many-body dynamics using subsystem {Loschmidt} echos},
  author = {Karch, Simon and Bandyopadhyay, Souvik and Sun, Zheng-Hang and Impertro, Alexander and Huh, SeungJung and Prieto Rodríguez, Irene and Wienand, Julian F. and Ketterle, Wolfgang and Heyl, Markus and Polkovnikov, Anatoli and Bloch, Immanuel and Aidelsburger, Monika},
  year   = {2025},
  note   = {\href{https://doi.org/10.48550/arXiv.2501.16995}{arXiv:2501.16995 [cond-mat.quant-gas]}}
}

@article{gkritsis_simulating_2024,
   title={Simulating {Chemistry} with {Fermionic} {Optical} {Superlattices}},
   volume={6},
   ISSN={2691-3399},
   doi={10.1103/PRXQuantum.6.010318},
   number={1},
   journal={PRX Quantum},
   publisher={American Physical Society (APS)},
   author={Gkritsis, Fotios and Dux, Daniel and Zhang, Jin and Jain, Naman and Gogolin, Christian and Preiss, Philipp M.},
   year={2025},
   month=jan,
   pages = {010318}
}

@article{bojovic_high-fidelity_2025,
  title = {High-fidelity collisional quantum gates with fermionic atoms},
  journal = {Nature},
  volume = {652},
  doi = {10.1038/s41586-026-10356-3},
  author = {Bojovi\'{c}, P. and Hilker, T. and Wang, S. and others},
  year = {2026},
  pages = {602--608}
}

@article{trebst_d-wave_2006,
	title = {$d$-wave resonating valence bond states of fermionic atoms in optical lattices},
	volume = {96},
	issn = {1079-7114},
	doi = {10.1103/PhysRevLett.96.250402},
	number = {25},
	journal = {Physical Review Letters},
	author = {Trebst, Simon and Schollwöck, Ulrich and Troyer, Matthias and Zoller, Peter},
	month = jun,
	year = {2006},
    pages = {250402}
}

@article{Rey_controlled_2009,
   title={Controlled preparation and detection of d-wave superfluidity in two-dimensional optical superlattices},
   volume={87},
   ISSN={1286-4854},
   url={http://dx.doi.org/10.1209/0295-5075/87/60001},
   DOI={10.1209/0295-5075/87/60001},
   number={6},
   journal={EPL (Europhysics Letters)},
   publisher={IOP Publishing},
   author={Rey, A. M. and Sensarma, R. and Fölling, S. and Greiner, M. and Demler, E. and Lukin, M. D.},
   year={2009},
   month=sep, pages={60001}
   }

@article{nascimbene_experimental_2012,
	title = {Experimental {Realization} of {Plaquette} {Resonating} {Valence}-{Bond} {States} with {Ultracold} {Atoms} in {Optical} {Superlattices}},
	volume = {108},
	url = {https://link.aps.org/doi/10.1103/PhysRevLett.108.205301},
	doi = {10.1103/PhysRevLett.108.205301},
	number = {20},
	urldate = {2025-09-01},
	journal = {Physical Review Letters},
	author = {Nascimbène, S. and Chen, Y.-A. and Atala, M. and Aidelsburger, M. and Trotzky, S. and Paredes, B. and Bloch, I.},
	month = may,
	year = {2012},
	pages = {205301},
}

@ARTICLE{Jordan_uber_1928,
       author = {{Jordan}, P. and {Wigner}, E.},
        title = "{{\"U}ber das {Paulische} {\"A}quivalenzverbot}",
      journal = {Zeitschrift fur Physik},
         year = 1928,
        month = sep,
       volume = {47},
       number = {9-10},
        pages = {631-651},
          doi = {10.1007/BF01331938},
       adsurl = {https://ui.adsabs.harvard.edu/abs/1928ZPhy...47..631J}
}

@article{Bravyi_fermionic_2002,
   title={Fermionic {Quantum} {Computation}},
   volume={298},
   ISSN={0003-4916},
   url={http://dx.doi.org/10.1006/aphy.2002.6254},
   DOI={10.1006/aphy.2002.6254},
   number={1},
   journal={Annals of Physics},
   publisher={Elsevier BV},
   author={Bravyi, Sergey B. and Kitaev, Alexei Yu.},
   year={2002},
   month=may, pages={210–226}
}

@article{Verstraete_mapping_2005,
   title={Mapping local {Hamiltonians} of fermions to local {Hamiltonians} of spins},
   volume={2005},
   ISSN={1742-5468},
   url={http://dx.doi.org/10.1088/1742-5468/2005/09/P09012},
   DOI={10.1088/1742-5468/2005/09/p09012},
   number={09},
   journal={Journal of Statistical Mechanics: Theory and Experiment},
   publisher={IOP Publishing},
   author={Verstraete, F and Cirac, J I},
   year={2005},
   month=sep, pages={P09012–P09012}
}

@article{Yu_clifford_2025,
   title={Clifford Circuit-Based Heuristic Optimization of Fermion-To-Qubit Mappings},
   volume={21},
   ISSN={1549-9626},
   doi={10.1021/acs.jctc.5c00794},
   number={19},
   journal={Journal of Chemical Theory and Computation},
   publisher={American Chemical Society (ACS)},
   author={Yu, Jeffery and Liu, Yuan and Sugiura, Sho and Van Voorhis, Troy and Zeytinoğlu, Sina},
   year={2025},
   month=sep,
   pages={9430–9443}
}

@article{Zhang_scalable_2023,
title={{Scalable} {Multipartite} {Entanglement} {Created} by {Spin} {Exchange} in an {Optical} {Lattice}},
   volume={131},
   ISSN={1079-7114},
   doi={10.1103/PhysRevLett.131.073401},
   number={7},
   journal={Physical Review Letters},
   publisher={American Physical Society (APS)},
   author={Zhang, Wei-Yong and He, Ming-Gen and Sun, Hui and Zheng, Yong-Guang and Liu, Ying and Luo, An and Wang, Han-Yi and Zhu, Zi-Hang and Qiu, Pei-Yue and Shen, Ying-Chao and Wang, Xuan-Kai and Lin, Wan and Yu, Song-Tao and Li, Bin-Chen and Xiao, Bo and Li, Meng-Da and Yang, Yu-Meng and Jiang, Xiao and Dai, Han-Ning and Zhou, You and Ma, Xiongfeng and Yuan, Zhen-Sheng and Pan, Jian-Wei},
   year={2023},
   month=aug,
   pages = {073401}
}

@article{xu_neutral-atom_2025,
   title={A neutral-atom {Hubbard} quantum simulator in the cryogenic regime},
   volume={642},
   ISSN={1476-4687},
   doi={10.1038/s41586-025-09112-w},
   journal={Nature},
   number={8069},
   publisher={Springer Science and Business Media LLC},
   author={Xu, Muqing and Kendrick, Lev Haldar and Kale, Anant and Gang, Youqi and Feng, Chunhan and Zhang, Shiwei and Young, Aaron W. and Lebrat, Martin and Greiner, Markus},
   year={2025},
   month=jun, pages={909–915}
}

@misc{zhang_observation_2025,
  title  = {{Observation} of {Coherent} {Quantum} {Tunneling} of a {Massive} {Atomic} {Cluster} with 435 u},
  author = {Zhang, Han and Wang, Yong-Kui and Zheng, Yi and Bai, Hai-Tao and Yang, Bing},
  year   = {2025},
  note   = {\href{https://doi.org/10.48550/arXiv.2502.06246}{arXiv:2502.06246 [cond-mat.quant-gas]}}
}

@article{Zhu_splitting_2025,
   title={{Splitting} and {Connecting} {Singlets} in {Atomic} {Quantum} {Circuits}},
   volume={15},
   ISSN={2160-3308},
   doi={10.1103/xh3v-tky4},
   number={4},
   journal={Physical Review X},
   publisher={American Physical Society (APS)},
   author={Zhu, Zijie and Kiefer, Yann and Jele, Samuel and Gächter, Marius and Bisson, Giacomo and Viebahn, Konrad and Esslinger, Tilman},
   year={2025},
   month=nov,
   pages = {041032}
}

@article{Corboz_improved_2016,
   title={Improved energy extrapolation with infinite projected entangled-pair states applied to the two-dimensional {Hubbard} model},
   volume={93},
   ISSN={2469-9969},
   doi={10.1103/PhysRevB.93.045116},
   number={4},
   journal={Physical Review B},
   publisher={American Physical Society (APS)},
   author={Corboz, Philippe},
   year={2016},
   month=jan,
   pages = {045116}
}

@misc{farhi_quantum_2000,
  title  = {Quantum {Computation} by {Adiabatic} {Evolution}},
  author = {Farhi, Edward and Goldstone, Jeffrey and Gutmann, Sam and Sipser, Michael},
  year   = {2000},
  note   = {\href{https://doi.org/10.48550/arXiv.quant-ph/0001106}{arXiv:quant-ph/0001106}}
}

@article{Bakr_a-quantum-gas_2009,
   title={A quantum gas microscope for detecting single atoms in a {Hubbard}-regime optical lattice},
   volume={462},
   ISSN={1476-4687},
   url={http://dx.doi.org/10.1038/nature08482},
   DOI={10.1038/nature08482},
   number={7269},
   journal={Nature},
   publisher={Springer Science and Business Media LLC},
   author={Bakr, Waseem S. and Gillen, Jonathon I. and Peng, Amy and Fölling, Simon and Greiner, Markus},
   year={2009},
   month=nov, pages={74–77}
}

@article{Parsons_site-resolved_2015,
   title={Site-{Resolved} {Imaging} of {Fermionic} $^6${Li} in an {Optical} {Lattice}},
   volume={114},
   ISSN={1079-7114},
   doi={10.1103/PhysRevLett.114.213002},
   number={21},
   journal={Physical Review Letters},
   publisher={American Physical Society (APS)},
   author={Parsons, Maxwell F. and Huber, Florian and Mazurenko, Anton and Chiu, Christie S. and Setiawan, Widagdo and Wooley-Brown, Katherine and Blatt, Sebastian and Greiner, Markus},
   year={2015},
   month=may,
   pages = {213002}
}

@article{Cheuk_quantum-gas_2015,
   title={Quantum-{Gas} {Microscope} for {Fermionic} {Atoms}},
   volume={114},
   ISSN={1079-7114},
   doi={10.1103/PhysRevLett.114.193001},
   number={19},
   journal={Physical Review Letters},
   publisher={American Physical Society (APS)},
   author={Cheuk, Lawrence W. and Nichols, Matthew A. and Okan, Melih and Gersdorf, Thomas and Ramasesh, Vinay V. and Bakr, Waseem S. and Lompe, Thomas and Zwierlein, Martin W.},
   year={2015},
   month=may,
   pages = {193001}
}

@article{khaneja_optimal_2005,
	title = {Optimal control of coupled spin dynamics: design of {NMR} pulse sequences by gradient ascent algorithms},
	volume = {172},
	issn = {1090-7807},
	doi = {10.1016/j.jmr.2004.11.004},
	number = {2},
	journal = {Journal of Magnetic Resonance},
	author = {Khaneja, Navin and Reiss, Timo and Kehlet, Cindie and Schulte-Herbrüggen, Thomas and Glaser, Steffen J.},
	year = {2005},
	pages = {296--305},
}

@article{childs_hamiltonian_2012,
author = {Childs, Andrew M. and Wiebe, Nathan},
title = {Hamiltonian simulation using linear combinations of unitary operations},
year = {2012},
issue_date = {November 2012},
publisher = {Rinton Press, Incorporated},
address = {Paramus, NJ},
volume = {12},
number = {11–12},
   doi={10.26421/QIC12.11-12-1},
issn = {1533-7146},
journal = {Quantum Info. Comput.},
month = nov,
pages = {901–924},
numpages = {24}
}

@article{Trivedi_quantum_2024,
  author    = {Rahul Trivedi and Adrian Franco Rubio and J. Ignacio Cirac},
  title     = {Quantum advantage and stability to errors in analogue quantum simulators},
  journal   = {Nature Communications},
  year      = {2024},
  volume    = {15},
  number    = {1},
  pages     = {6507},
  doi       = {10.1038/s41467-024-50750-x},
  issn      = {2041-1723}
}

@article{schollwock_density-matrix_2011,
	title={The density-matrix renormalization group in the age of matrix product states},
   volume={326},
   ISSN={0003-4916},
   DOI={10.1016/j.aop.2010.09.012},
   number={1},
   journal={Annals of Physics},
   publisher={Elsevier BV},
   author={Schollwöck, Ulrich},
   year={2011},
   month=jan,
   pages={96–192}
   }

@article{fishman_itensor_2022,
title={The {ITensor} {Software} {Library} for {Tensor} {Network} {Calculations}},
   doi={10.21468/SciPostPhysCodeb.4},
   journal={SciPost Physics Codebases},
   publisher={Stichting SciPost},
   author={Fishman, Matthew and White, Steven and Stoudenmire, Edwin},
   year={2022},
   month=aug }

@article{schiffer_virtual_2024,
	title = {Virtual mitigation of coherent non-adiabatic transitions by echo verification},
	volume = {8},
	doi = {10.22331/q-2024-05-14-1346},
	urldate = {2025-08-21},
	journal = {Quantum},
	author = {Schiffer, Benjamin F. and Vreumingen, Dyon van and Tura, Jordi and Polla, Stefano},
	month = may,
	year = {2024},
	pages = {1346},
}

@article{byrd_limited_1995,
	title = {A {Limited} {Memory} {Algorithm} for {Bound} {Constrained} {Optimization}},
	volume = {16},
	doi = {10.1137/0916069},
	number = {5},
	journal = {SIAM Journal on Scientific Computing},
	author = {Byrd, Richard H. and Lu, Peihuang and Nocedal, Jorge and Zhu, Ciyou},
	year = {1995},
	pages = {1190-1208},
}

@article{zhu_algorithm_1997,
author = {Zhu, Ciyou and Byrd, Richard H. and Lu, Peihuang and Nocedal, Jorge},
title = {Algorithm 778: {L-BFGS-B}: {F}ortran subroutines for large-scale bound-constrained optimization},
year = {1997},
issue_date = {Dec. 1997},
publisher = {Association for Computing Machinery},
address = {New York, NY, USA},
volume = {23},
number = {4},
issn = {0098-3500},
doi = {10.1145/279232.279236},
journal = {ACM Trans. Math. Softw.},
month = dec,
pages = {550–560},
numpages = {11}
}

@article{Weinberg_quspin_2017,
   title={{QuSpin}: a {Python} package for dynamics and exact diagonalisation of quantum many body systems. Part {I}: spin chains},
   volume={2},
   doi={10.21468/SciPostPhys.2.1.003},
   number={1},
   journal={SciPost Phys.},
   author={Weinberg, Phillip and Bukov, Marin},
   year={2017},
   pages={003}
}

@article{Weinberg_quspin_2019,
   title={{QuSpin}: a {Python} package for dynamics and exact diagonalisation of quantum many body systems. Part {II}: bosons, fermions and higher spins},
   volume={7},
   doi={10.21468/SciPostPhys.7.2.020},
   number={2},
   journal={SciPost Phys.},
   author={Weinberg, Phillip and Bukov, Marin},
   year={2019},
   pages={020}
}

@article{kiefer_protected_2025,
    title = {Protected quantum gates using qubit doublons in dynamical optical lattices},
    volume = {652},
    issn = {1476-4687},
    doi = {10.1038/s41586-026-10285-1},
    number = {8052},
    journal = {Nature},
    author = {Kiefer, Yann and Zhu, Zijie and Fischer, Lars and Jele, Samuel and G{\"a}chter, Marius and Bisson, Giacomo and Viebahn, Konrad and Esslinger, Tilman},
    year = {2026},
    pages = {609--614},
}

@article{Klemmer_floquet-driven_2024,
   title={{Floquet}-{Driven} {Crossover} from {Density}-{Assisted} {Tunneling} to {Enhanced} {Pair} {Tunneling}},
   volume={133},
   ISSN={1079-7114},
   doi={10.1103/PhysRevLett.133.253402},
   number={25},
   journal={Physical Review Letters},
   publisher={American Physical Society (APS)},
   author={Klemmer, Nick and Fleper, Janek and Jonas, Valentin and Sheikhan, Ameneh and Kollath, Corinna and Köhl, Michael and Bergschneider, Andrea},
   year={2024},
   month=dec,
   pages = {253402}
}

@misc{ITEcode,
  author       = {Alberto R. Cavallar and Benjamin F. Schiffer},
  title        = {{ITE} {Ramp} {Optimization} {Code} for {Phase}-{Sensitive} {Measurements} on a {Fermi}-{Hubbard} {Quantum} {Processor}},
  note         = {{GitLab} repository, version v1.0-paper, \url{https://gitlab.physik.uni-muenchen.de/A.Cavallar/ite-ramp-optimization.git}},
  year         = {2025}
}

\appendix

\section{Details for adiabatically prepared plaquette states}

\subsection{Adiabatic preparation of plaquette states}
\label{appendix:adiabatic_prep}

We outline the preparation of the plaquette ground state, which is the first step of the experimental protocol.
Starting from two isolated double wells, each double well in the $|\updownarrows,0\rangle$ Fock state, the plaquette ground state can be prepared by a two-step, piecewise-linear adiabatic path --- in the spirit of earlier proposals~\cite{trebst_d-wave_2006}.

The initial Hamiltonian on the plaquette is
\begin{align}
    H_0=H_\text{FH}^{2\times 2}(t_x\!=\!0,t_y\!=\!0,U\!=\!0,\Delta \mu_x \!=\! t),
\end{align}
where a horizontal tilt of the chemical potential $\Delta\mu_x$ energetically favors the doubly occupied wells.
The ground state $\ket{\psi_0}=\ket{\updownarrows,0}^{\otimes 2}$ is then delocalized by ramping $\Delta\mu_x$ to zero, while simultaneously increasing the horizontal hoppings $t_x$.
This first step prepares the intermediate ground state
\begin{align}
    |\psi_\text{i}\rangle=\big[\tfrac{1}{2}(\ket{\uparrow,\downarrow}-\ket{\downarrow,\uparrow}+\ket{\updownarrows,0} + \ket{0,\updownarrows})\big]^{\otimes2}
\end{align}
of the Hamiltonian
\begin{align}
    H_\text{i}=H_{\text{FH}}^{2\times 2}(t_x\!=\!t, t_y\!=\!0,U\!=\!0,\Delta\mu_x\!=\!0).
\end{align}
Alternatively, this step may be realized by a digital gate sequence (cf.~Ref.~\cite{mark_efficiently_2024}).
The plaquette ground state $\ket{\psi_1}$ is then obtained by adiabatically ramping up both the vertical hoppings $t_y$ and the on-site interactions $U$, reaching the target Hamiltonian
\begin{align}
    H_{1} =H_{\text{FH}}^{2\times 2}(t_x\!=\!t, t_y\!=\!t, U\!=\!8\,t,\Delta\mu_x\!=\!0).
\end{align}
Together, the adiabatic path can be written as an interpolation in parametrized time $s=\tau/\tau_\text{total}$ from $s=0$ to $s=1$ in a piecewise-linear manner as
\begin{align}
	H(s)
	 & = [(1-2s)H_0 + 2sH_\text{i}]\theta(1/2-s)\nonumber   \\
	 & \quad + [(2-2s)H_\text{i} + (2s-1)H_1]\theta(s-1/2),
\end{align}
with $\theta(s)$ the Heaviside step function.
The plaquette ground state can be prepared adiabatically because the adiabatic path remains gapped throughout the evolution with these choices of $H_0$ and $H_\text{i}$.

\subsection{Numerical plaquette ground state energies} \label{appendix:TN_data}

\begin{figure}[tb]
    \centering
    \includegraphics[width=.9\linewidth]{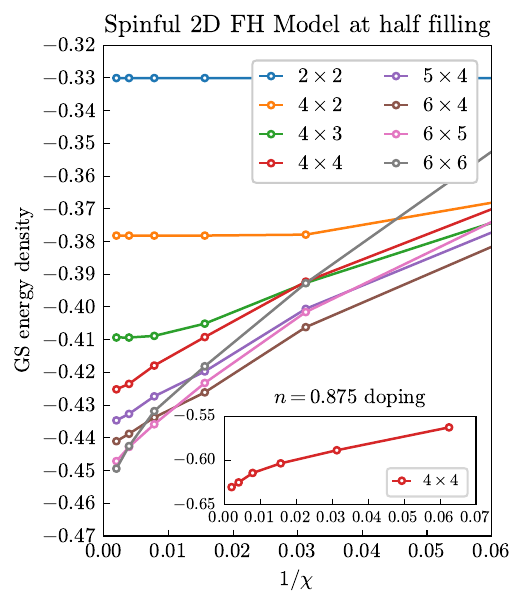}
    \caption{
    \justifying
    Ground state energy density of the Fermi--Hubbard model at different system sizes at half-filling. 
    \textbf{(Inset)} Doped $4\times4$ system at $n=0.875=1-1/8$ doping.}
    \label{fig:app_dmrg_halffilled}
\end{figure}

We include the data used for the estimation of the plaquette ground state energy density for different plaquette sizes. 
We perform a variational ground state search using DMRG~\cite{schollwock_density-matrix_2011} on matrix-product states in a snake-like geometry for different values of the maximum bond dimension $\chi$. 
The simulations are implemented using the \texttt{ITensor} package~\cite{fishman_itensor_2022}.
In Fig.~\ref{fig:app_dmrg_halffilled}, we show the ground state energy as a function of the inverse bond dimension both for the half-filled case at different system sizes and for the doped case ($n=0.875$) on a lattice of $4\times 4$ sites.
The extrapolation to vanishing $1/\chi$ is performed manually with a corresponding uncertainty and included in the main text in Fig.~\ref{fig:fig2}(c).

\subsection{Projective measurement for adiabatically prepared states}
\label{appendix:proj_adiabatic}

We discuss how to undo the adiabatic preparation, which is required for estimating the amplitude of the Loschmidt echo through projective Fock-basis measurements (cf.~Sec.~\ref{sec:implementation}).

We denote the (perfectly) adiabatic state preparation as the unitary $\mathcal{U}_{0\rightarrow 1}$, mapping the ground state $\ket{\psi_0}$ of an initial Hamiltonian $H_0$, to the target ground state $\ket{\psi_1}$ of a final Hamiltonian $H_1$ via the adiabatic path $\widetilde{H}(s)=(1-s)H_0 + sH_1$ with a non-vanishing spectral gap between the ground state and the first excited state for any $s\in[0,1]$.
Explicitly, the unitary is defined as
\begin{align}
    \mathcal{U}_{0\rightarrow 1} \coloneqq\mathcal{T} \exp\Big(-i\tau_\text{total}\int_0^1\mathrm{d}s\,\widetilde{H}(s)\Big),
\end{align}
with $\mathcal{T}$ the time-ordering operator.
Here, the single plaquette ground state $\ket{\psi_1}$ is adiabatically prepared from the Fock state $\ket{\psi_0}=\ket{\updownarrows,0}^{\otimes 2}$.

Our protocol requires projective measurements onto the initial plaquette state $\ket{\Psi_p}=\otimes_\mu \ket{\psi_1}_\mu$ to determine Loschmidt-echo overlaps with the time-evolved state, i.e.,~$\langle\Psi_p|\exp(-iH_\text{FH}\tau)|\Psi_p\rangle$.
However, only Fock-basis measurements are available with a quantum gas microscope. 
One therefore needs in principle to uncompute the adiabatic preparation as the inverse $\mathcal{U}^\dagger_{0\rightarrow 1}$ and only then project onto the Fock state basis, counting the number of occurrences of the Fock state $\ket{\psi_0}$.
This could be achieved by evolving backwards in time, yet this is typically not available on quantum simulators.
Instead, it is sufficient to reverse the path direction as
\begin{equation}
    \mathcal{U}_{1\rightarrow 0}\coloneqq\mathcal{T} \exp \Big(-i\tau_\text{total}\int_1^0\mathrm{d}s\,\widetilde{H}(s)\Big).
\end{equation}
This operation maps the final ground state back to the initial ground state, although the dynamics for excited states is generally more complex.
See also~\cite{schiffer_virtual_2024} for a related error mitigation strategy.
We note that the global phase imprinted on the ground state will be different, i.e.,
\begin{align}
    \mathcal{U}_{1\rightarrow 0}\ket{\psi_1}=e^{i\varphi}\ket{\psi_0}\neq \ket{\psi_0}=
\mathcal{U}_{0\rightarrow 1}^\dagger\ket{\psi_1}.
\end{align}
However, this is not problematic, since a global phase will not be detectable in the projective measurement.

\section{Details for the ITE implementation on the initial plaquette state}

\subsection{Details for tiling the ITE unitaries}
\label{appendix:ITE_scalability}

In Sec.~\ref{sec:implementation} we describe how the unitary $V$ implementing the (normalized) ITE $\exp(-H_\Lambda \vartheta)$, with $H_\Lambda = \sum_m H_m$ can be decomposed into local unitaries $V_m$ as defined in Eq.~\eqref{ITEmpsi}.
We now show why these unitaries can be defined acting on the \textit{same} initial state $\ket{\Psi}$ instead of acting each on the evolved state $|\Psi_{m-1}\rangle\coloneqq V_{m-1}\cdots V_1|\Psi\rangle$ after the previous layer, as long as the imaginary time $\vartheta$ is small.

Let $\{H_m\}_{m=1}^4$ be the Hamiltonians that involve only the hopping terms $h_{\mu\nu}$ coupling plaquettes and distributed in parallel as in Fig.~\ref{fig:fig2}(f), respectively corresponding to unitaries $\{V_m\}_{m=1}^4$ implementing the ITE up to normalization:
\begin{align}
    V_m |\Psi\rangle &= \frac{1}{\mathcal{N}_m} e^{-H_m\vartheta} |\Psi\rangle \nonumber\\
    &= \frac{1}{\mathcal{N}_m} [\mathds{1}- H_m\vartheta + \mathcal{O}(\vartheta^2)]|\Psi\rangle.
\end{align}
The normalization $\mathcal{N}$ is given as
\begin{align}
    \mathcal{N}_m^2 &= \langle\Psi| e^{-2H_m\vartheta} |\Psi\rangle \nonumber\\
    &= 1-2\vartheta\langle\Psi| H_m |\Psi\rangle + \mathcal{O}(\vartheta^2),
\end{align}
such that
\begin{align}
    1/\mathcal{N}_m = 1 + \vartheta\langle\Psi| H_m |\Psi\rangle + \mathcal{O}(\vartheta^2).
\end{align}
Considering the full unitary sequence, we obtain the expression
\begin{align}
    &\textstyle\prod_m V_m |\Psi\rangle  \nonumber\\
    =& \frac{1}{\prod_m \mathcal{N}_m} \big[\mathds{1} - \vartheta \textstyle\sum_m H_m + \mathcal{O}(\vartheta^2)\big]|\Psi\rangle\nonumber\\
     = & \big[ \mathds{1} + \vartheta\textstyle\sum_m \big(\langle\Psi|H_m |\Psi\rangle - H_m\big) + \mathcal{O}(\vartheta^2)\big]|\Psi\rangle\nonumber\\
     =&\left[\mathds{1}+\vartheta(\langle\Psi|H_\Lambda |\Psi\rangle -  H_\Lambda) \right]|\Psi\rangle+\mathcal{O}(\vartheta^2).
\end{align}
This matches the target state after short ITE
\begin{align}
    & \frac{1}{\mathcal{N}} e^{-H_\Lambda\vartheta} |\Psi\rangle\nonumber\\
    & =\left[\mathds{1}+\vartheta(\langle\Psi|H_\Lambda |\Psi\rangle -  H_\Lambda)\right]|\Psi\rangle + \mathcal{O}(\vartheta^2),
\end{align}
thus showing that the error is
\begin{align}
    \textstyle\big|\big|\big(V-\prod_m V_m\big)|\Psi\rangle\big|\big| = \mathcal{O}(\vartheta^2),
\end{align}
i.e., of the same order in $\vartheta$ as the Trotterization error.

Note that such an error scaling will be achieved by any permutation of the $\{V_m\}_{m=1}^4$ sequence, even though the first unitary in the sequence will lead to a small amount of entanglement between the initially unentangled plaquettes.
Importantly, this enables the scalability of our approach in the sense that a single building block unitary $V_{\mu\nu}$ --- given by the pulse sequence in Fig.~\ref{fig:fig3} and composing the layers $V_m$ as in Fig.~\ref{fig:fig2}(f) --- is sufficient to implement the ITE up to normalization on an arbitrarily large lattice.

\subsection{Details on numerical pulse optimization}
\label{appendix:ITE_pulse_optim}

In Sec.~\ref{sec:implementation} we describe how a pulse sequence modulating the hopping strengths of the Fermi–Hubbard Hamiltonian on the subsystem of plaquettes $P_\mu$ and $P_\nu$ can realize the (normalized) ITE unitary $V_{\mu\nu}$.
Here, we describe how the unitary is obtained with an optimal control approach in the spirit of the GRAPE protocol~\cite{khaneja_optimal_2005}.

The procedure consists of maximizing the overlap squared between the (normalized) imaginary-time evolved state~$\ket\Phi \propto \exp(-h_{\mu\nu} \vartheta)\ket{\Psi_p}$ and the state prepared by unitary dynamics 
\begin{align}
    \mathcal{T}\exp\Big(-i\int_0^\tau\mathrm{d}\tau'\, h(\tau') \Big) \ket{\Psi_p}    
\end{align}
under the time-dependent Hamiltonian $h(\tau)=H_\text{FH}(\tau)\big|_{\mu\nu}$, where the subscript implies a local Hamiltonian that is acting only onto these two plaquettes.
The optimization is achieved by discretizing the evolution into a finite number of steps with fixed step size $\Delta \tau$ and optimizing over the parameters $\mathbf{p}=(\mathbf{p}_1,...,\mathbf{p}_\ell)$, such that
\begin{align}
    \max_\mathbf{p} | \langle\Phi| e^{-i h(\mathbf{p}_\ell)\Delta \tau} \cdots e^{-i h(\mathbf{p}_1)\Delta \tau} |\Psi_p\rangle |^2.
\end{align}
Concretely, we consider as parameters $\mathbf{p}$ the different time-dependent tunneling strengths.
Due to symmetry arguments, the three free independent hopping strengths are $t_x,t_x',t_y$ in the superlattice architecture, as depicted in Fig.~\ref{fig:fig3}(a).
The ITE pulse is parametrized pointwise in time to maximize flexibility, with unfixed endpoints.
While optimizations over smooth basis functions are possible and may yield experimentally smoother pulses, we do not impose such a restriction here, as the optimal ramp shape is not known a priori.

The numerical optimization is performed using Python's \texttt{scipy.optimize.minimize} with the bound-constrained minimization method L-BFGS-B \cite{byrd_limited_1995, zhu_algorithm_1997} guaranteeing that the hopping control parameters remain positive and bounded, $0\leq t_i/t\leq 10$, during the optimization, with a step size of $\Delta \tau = 0.05/t$.
The source code together with the data corresponding to the found optimal pulses are publicly available~\cite{ITEcode}.

\subsection{ITE on a double well}

\begin{figure}[t]
    \centering
    \includegraphics[width=0.9\linewidth]{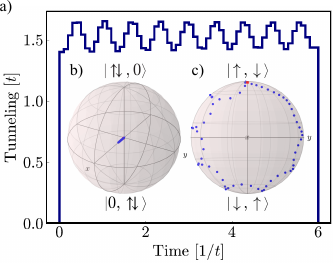}
    \caption{
    \justifying
    \textbf{(a)} ITE pulse sequence for a double well initialized in the $\ket{\uparrow, \downarrow}$ state, realized with 99.96\% fidelity via modulation of the tunneling amplitude between both wells while keeping $U=8t$ fixed.
    \textbf{(b-c)} Bloch sphere projections of the ITE gate dynamics onto the doublon and Néel state bases.
    }
    \label{fig:doublewell}
\end{figure}

We include in Fig.~\ref{fig:doublewell}(a) a simple example of ITE on a double well initialized in the $\ket{\uparrow,\downarrow}$ state.
We visualize the dynamics by projecting onto effective Bloch spheres as in Figs.~\ref{fig:doublewell}(b--c) for the doublon and N\'eel bases, respectively.
We observe that modulating the tunneling amplitude drives the initial state around these projected spheres before reaching the normalized ITE target state.

For intuition, note that for small $\vartheta$ one has
\begin{align}
	e^{-\vartheta H_\mathrm{FH}}\ket{\uparrow,\downarrow}
	= \ket{\uparrow,\downarrow} - \vartheta H_\mathrm{FH}\ket{\uparrow,\downarrow} + \mathcal{O}(\vartheta^2).
\end{align}
In the two-site Hubbard model, the leading correction to $\ket{\uparrow,\downarrow}$ comes from single hopping events that create doublon--holon configurations, i.e.,~it is of order $\vartheta t$ and lies in the subspace spanned by $\ket{\updownarrows,0}$ and $\ket{0,\updownarrows}$.
By contrast, a spin-swap component $\ket{\downarrow,\uparrow}$ arises only at second order through virtual doublon processes (superexchange).

\section{Details for doped plaquette states}

\subsection{Fock state contributions to doped ITE}
\label{appendix:Fock_doped}

\begin{figure}[b]
    \centering
    \includegraphics[width=\linewidth]{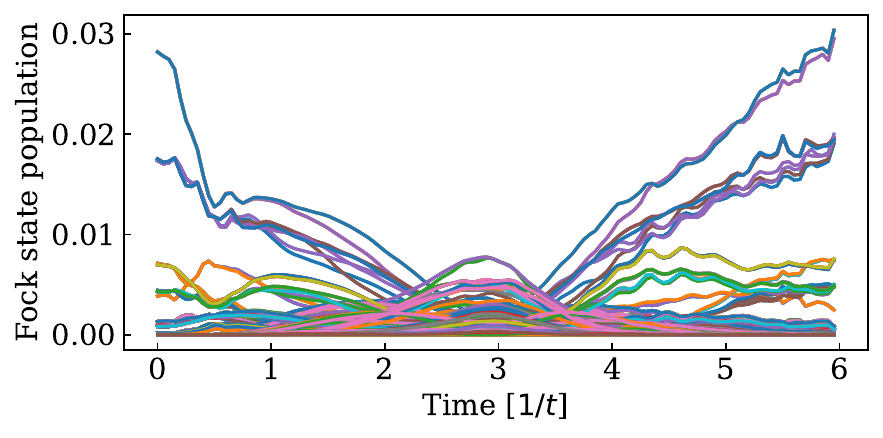}
    \caption{
    \justifying
    Instantaneous Fock state population throughout the ITE pulse sequence on the $AB$ doped state, revealing no clear dominant contributions.}
    \label{fig:fock_doped}
\end{figure}

An ITE pulse sequence for 12.5\%-doped plaquette states was presented in Sec.~\ref{sec:doping} [cf.~Fig.~\ref{fig:fig5}(e)].
We include the Fock state populations throughout the evolution in Fig.~\ref{fig:fock_doped}.
Unlike the half-filled case in Fig.~\ref{fig:fig3}(b), there do not seem to be clear dominant contributions from specific Fock states for the considered doped state.

\subsection{Alternative for doped plaquette states}
\label{appendix:alternative_doping}

\begin{figure*}
    \centering 
    \includegraphics[width=0.85\linewidth]{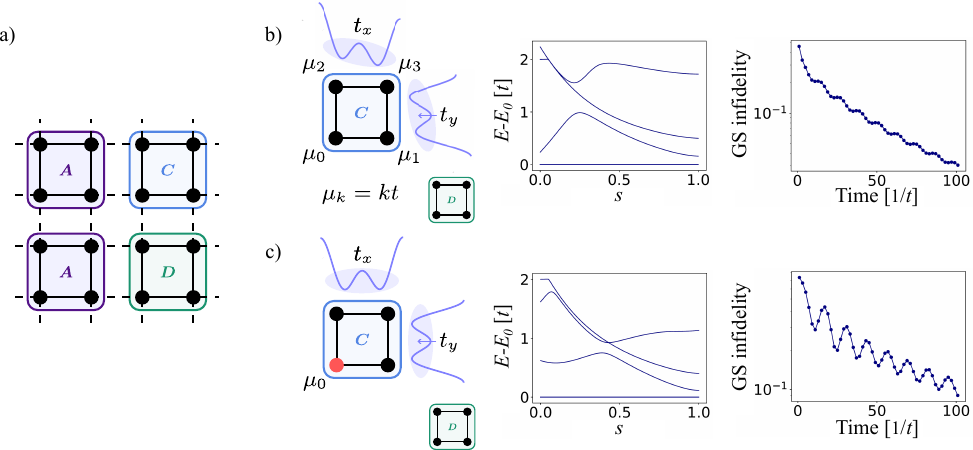}
    \caption{
    \justifying
    Adiabatic preparation of doped plaquettes.
    \textbf{(a)} A doping of 12.5\% may also be obtained from the plaquette ansatz by replacing every second half-filled plaquette $A=(2\uparrow,2\downarrow)$ by alternating the doped ones $C=(2\uparrow,1\downarrow)$ and $D=(1\uparrow,2\downarrow)$, giving rise to an $ACAD$ unit cell.
    Each $C$ (resp. $D$) plaquette ground state can be prepared adiabatically from a doublon Fock state $|\updownarrows,0\rangle$ and a $\ket{\uparrow,0}$ (resp. $\ket{\downarrow,0}$) state, each on a horizontal double well.
    \textbf{(b)} First,
    each Fock state is delocalized to the ground state within each double well.
    Then, the plaquette state is reached by slowly increasing the hopping along the vertical direction $t_y\rightarrow t$, jointly with the on-site interaction $U\rightarrow 8\,t$, while leaving $t_x =t$ fixed in the presence of a potential gradient $\mu_k$ breaking the existing ground state degeneracy.
    High-fidelity state preparation requires longer sweep times than for the variant presented in the main text.
    \textbf{(c)} The energy gap may alternatively be opened by instead adding a local energy offset $\mu_0$.}
    \label{fig:fig8}
\end{figure*}

We briefly discuss an alternative way to construct an initial plaquette state for 12.5\% doping.
Instead of combining three half-filled plaquettes~$A$ with a doped plaquette $B$ that has two spins removed, we can consider two $A$ plaquettes and a doped plaquette $C$ missing a spin-down particle together with a plaquette $D$ where one spin-up is removed, as illustrated in Fig.~\ref{fig:fig8}(a).
This $ACAD$ combination has a ground state energy density of $e/t=-0.46$, which is higher than in the $AAAB$ configuration. 
As this alternative is likely more challenging to realize experimentally, we focus on the $AAAB$ structure in the main text.

We further note that the plaquettes $C$ and $D$ can both be prepared through the same adiabatic ramp due to their spin symmetry.
In Figs.~\ref{fig:fig8}(b-c) we include two options for the adiabatic preparation, both requiring an additional potential as before, breaking the ground state degeneracy.
The first option is identical to the adiabatic preparation of $B$ plaquettes as discussed in the main text.
The second option consists of using the local potential $\mu_k = \delta_{k,0}t$ instead of the potential gradient $\mu_k = kt$.
The former could be realized by local addressing with a laser beam.

\section{Details on numerical sample complexity analysis} \label{appendix:numerical_samples}

Here we include an additional analysis on the LDOS reconstruction from Sec.~\ref{sec:benchmark} and provide a few technical details for completeness.

\begin{figure*}[p]
    \centering
    \includegraphics[width=\linewidth]{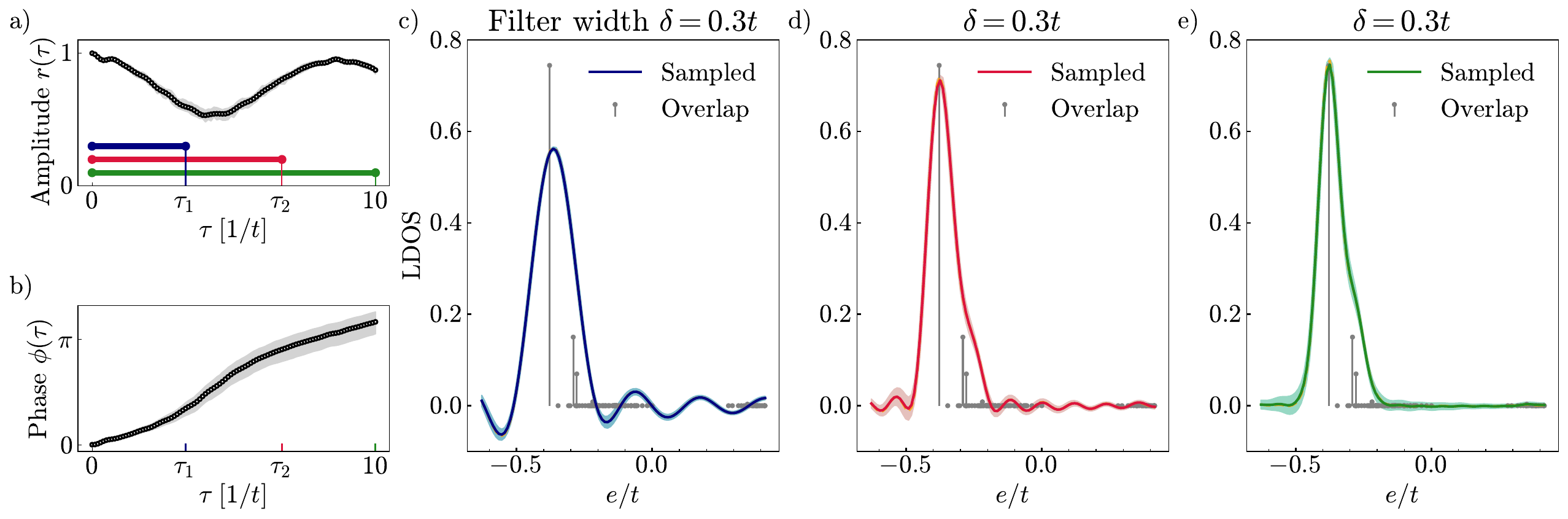}
    \caption{
    \justifying
    Shot noise analysis of the broadened local density of states (LDOS) for a \textit{fixed} filter width. The amplitude \textbf{(a)} and phase \textbf{(b)} of the Loschmidt echoes are computed as a function of time for the Fermi–Hubbard Hamiltonian at $U/t=8$ on a $4\times 2$ lattice at half-filling.
    The mean energy contribution to the phase has been subtracted for better clarity.
    \textbf{(c-e)} For a fixed Gaussian filter width $\delta=0.3\,t$, Loschmidt echoes up to times $\tau_1=\tau_3/3$, $\tau_2=2\tau_3/3$, $\tau_3=10\,[1/t]$ are respectively used to estimate the LDOS.
    Each of the amplitudes $r(\tau)$ and $r(\tau\pm i\vartheta)$ has been estimated with 100 samples.
    The mean value is shown as a solid curve with a shaded region showing one standard deviation of the data.
    The exact curve resulting from the ITE algorithm without sampling is shown as a dashed line in yellow, hardly visible due to the high precision of the overlying sampled curves.
    Gray bars with markers indicate overlaps of
    $\ket{\Psi_p}$ with the $4\times 2$ Fermi–Hubbard eigenstates.
    }
    \label{fig:app_LDOS}
\end{figure*}

We probe the effect of the finite time cutoff in the approximation of the Fourier transformation. 
We do so by fixing the Gaussian filter width to $\delta=0.3\,t$. 
Fig.~\ref{fig:app_LDOS} shows that the unphysical density of states below zero is indeed attributed to numerical errors in the filter's truncation, as the wiggling of both the sampled (solid) and dashed curves is strongly reduced when increasing the maximum time~$\tau_{\max}$.
We note that the LDOS reliably identifies the ground state energy already for the shorter maximal time $\tau_2$ (red).

The Loschmidt echoes in the numerical simulations in Fig.~\ref{fig:fig4} and Fig.~\ref{fig:app_LDOS} are estimated by assuming an ideal state preparation and perfect ITE pulses using the \texttt{QuSpin} package \cite{Weinberg_quspin_2017,Weinberg_quspin_2019}.
Amplitudes are sampled from a binomial distribution with probabilities given by the exact values squared.
For clarity of presentation, in Fig.~\ref{fig:fig4}(b) and Fig.~\ref{fig:app_LDOS}(b) we have subtracted the phase contribution $\exp(-iE_p\tau)$ of the mean energy $E_p=\bra{\Psi_p}H_\text{FH}\ket{\Psi_p}$ of the initial state.
The Gaussian filter coefficients
\begin{align}
    c_m = \Delta \tau \frac{\delta}{\sqrt{2\pi}} \exp(-\tau_m^2 \delta^2/2)
\end{align}
are computed classically for discrete times labeled as $\tau_m = m\Delta \tau \in [0,\tau_{\max} ]$, i.e.~$m\in\{0,..., R=\lfloor \tau_{\max}/\Delta\tau\rfloor \}$.

\begin{figure*}[p]
    \centering
    \includegraphics[width=0.8\textwidth]{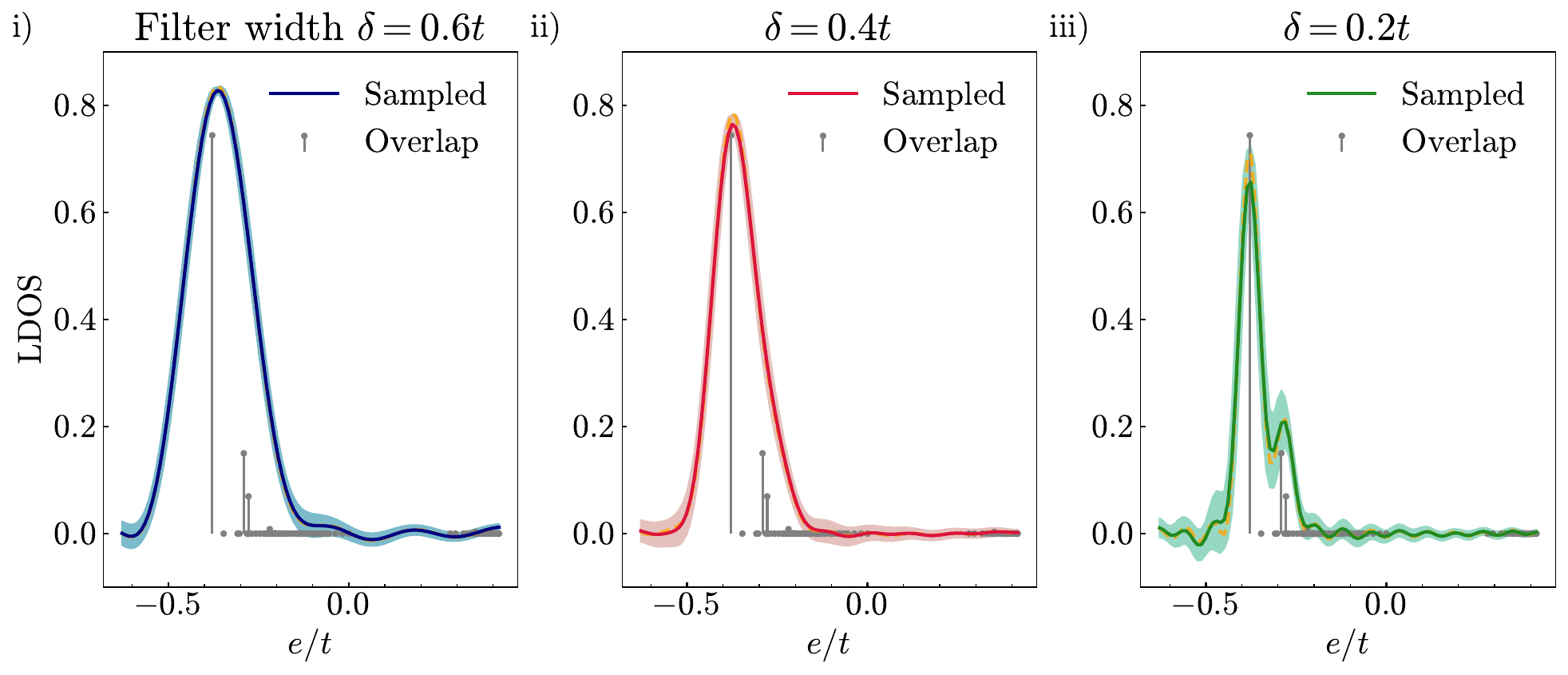}
    \includegraphics[width=0.8\textwidth]{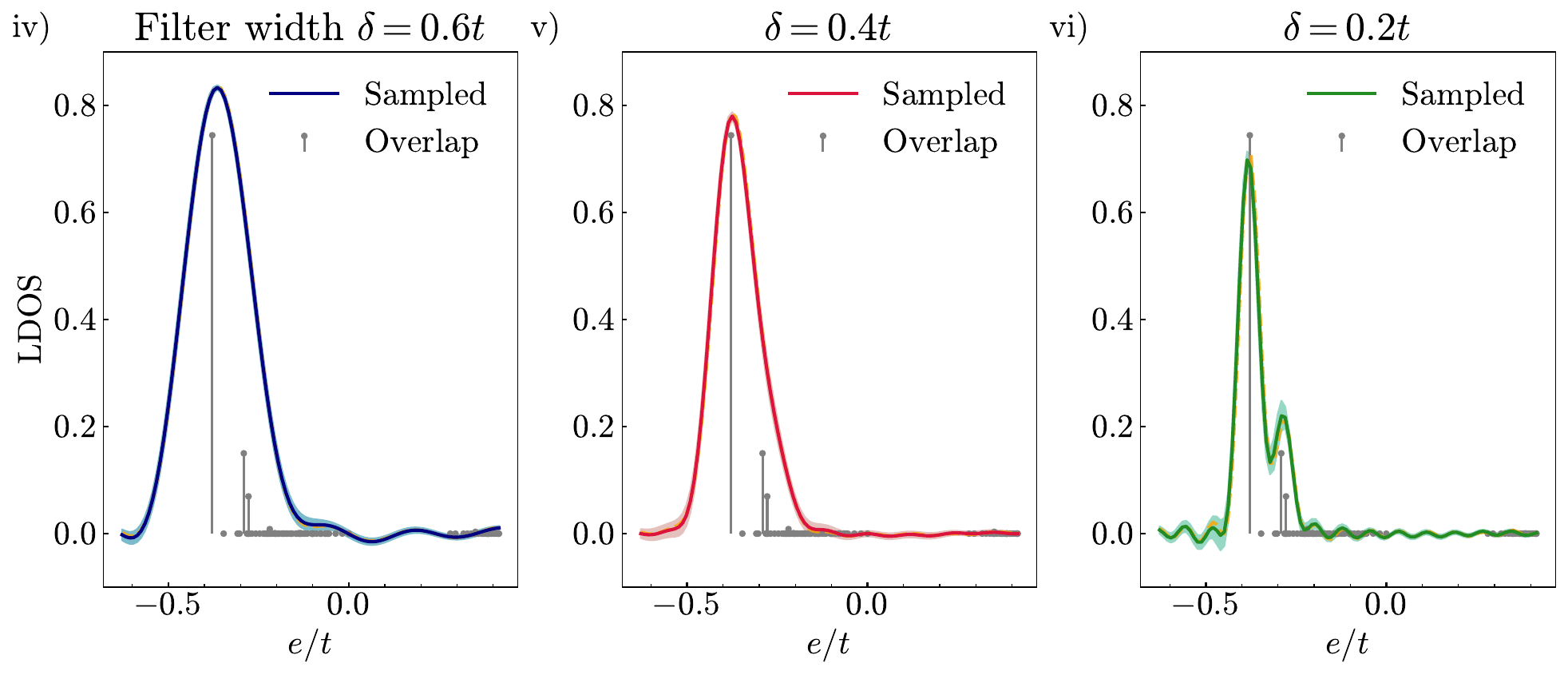}
    \caption{
    \justifying
    Influence of the short imaginary time $\vartheta$ on the accuracy of local density of states (LDOS) estimation.
    For Loschmidt echoes estimated from the ITE algorithm with short imaginary time \textbf{(i--iii)} $\vartheta=0.05/t$ and \textbf{(iv--vi)} $\vartheta = 0.2/t$, the respective LDOS are estimated with Gaussian filter widths $\delta=0.6t,0.4t,0.2t$ corresponding to total times $\tau_1=\tau_3/3$, $\tau_2=2\tau_3/3$, $\tau_3=10$ in units of $1/t$.
    Each of the amplitudes $r(\tau)$ and $r(\tau\pm i\vartheta)$ has been estimated with 100 samples.
    The mean value is shown as a solid curve with a shaded region indicating one standard deviation.
    The exact curve resulting from the ITE algorithm without sampling is shown as a dashed yellow line, hardly visible due to the high precision of the overlying sampled curves.
    Gray bars with markers indicate overlaps of the initial plaquette state $\ket{\Psi_p}$ with the $4\times2$ eigenstates of the Fermi--Hubbard Hamiltonian at $U/t=8$ and half filling.
    }
\label{fig:LDOS_thetas}
\end{figure*}

We have additionally analyzed the dependence of the local density of states (LDOS) estimation on the choice of the short imaginary time parameter $\vartheta$ entering the imaginary-time evolution (ITE) protocol.
In the main text we use $\vartheta=0.1/t$, a value commonly employed in the literature.
To assess robustness, we repeated the analysis by halving and doubling this value, shown in Figs.~\ref{fig:LDOS_thetas}.
We find that for significantly smaller $\vartheta$, the LDOS reconstruction becomes noticeably noisier, while also requiring more samples.
This can be understood from the fact that for very short imaginary times the ITE has little effect on the state, so that the differences between the shifted amplitudes $r(\tau\pm i\vartheta)$ become small and increasingly susceptible to statistical noise.
Doubling $\vartheta$, on the other hand, yields results of comparable accuracy but does not lead to a clear improvement.
Overall, this indicates that $\vartheta=0.1/t$ already provides a good compromise between signal strength and robustness. While somewhat larger values of $\vartheta$ could in principle be used, they are ultimately limited by the validity of the underlying approximations and by increasing Trotterization errors, which rely on $\vartheta$ remaining sufficiently small.

\end{document}